\documentclass[11pt,a4paper]{article}

\usepackage{jcappub,amsfonts,amsmath,amssymb,bm,graphicx}

\usepackage[active]{srcltx}

\title{Instability of magnetic fields in electroweak plasma driven by neutrino asymmetries}

\author[a,b,c]{Maxim Dvornikov}
\author[c]{Victor B. Semikoz}
\affiliation[a]{Research School of Physics and Engineering, Australian National University, 2601 Canberra, ACT, Australia}
\affiliation[b]{Institute of Physics, University of S\~{a}o Paulo, CP 66318, CEP 05315-970 S\~{a}o Paulo, SP, Brazil}
\affiliation[c]{Pushkov Institute of Terrestrial Magnetism, Ionosphere
and Radiowave Propagation of the Russian Academy of Sciences (IZMIRAN),
142190 Troitsk, Moscow, Russia}
\emailAdd{maxim.dvornikov@anu.edu.au}
\emailAdd{semikoz@yandex.ru}

\abstract{
The magnetohydrodynamics (MHD) is modified to incorporate the parity violation in the Standard Model
leading to a new instability of magnetic fields in the electroweak plasma in the presence of
nonzero neutrino asymmetries. The main ingredient for such a modified MHD is the antisymmetric part of
the photon polarization tensor in plasma, where the parity violating neutrino interaction with charged leptons
is present. We calculate this contribution to the polarization
tensor connected with the Chern-Simons term in effective Lagrangian of
the electromagnetic field. The general expression for such a contribution which depends on
the temperature and the chemical potential of plasma as well as on the photon's momentum is derived.
The instability of a magnetic field driven by the electron neutrino asymmetry for the $\nu$-burst during the first second of a supernova explosion can amplify a seed magnetic field of a protostar, and, perhaps, can explain the generation of strongest magnetic fields in magnetars.
The growth of a cosmological magnetic field driven by the neutrino asymmetry density $\Delta n_{\nu}=n_{\nu} - n_{\bar{\nu}}\neq 0$  is provided by a lower bound on $|\xi_{\nu_e}|=|\mu_{\nu_e}|/T$ which is consistent with the well-known Big Bang nucleosynthesis (upper) bound on neutrino asymmetries in a hot universe plasma.
}


\begin{document}

\maketitle

\section{Introduction}
The generation of the cosmological magnetic field (CMF) as a seed of observable galactic magnetic fields
is still an open problem \cite{Grasso:2000wj}. The two facts enhanced a new interest to such a problem. The first observational indications
of the presence of CMF in intergalactic medium which may survive even till the present epoch \cite{Neronov:2009gh,Neronov:1900zz} were as a new incitement to
the conception of CMF and its helicity. Secondly, there appeared some new models of the magnetic field instability leading to the generation of CMF. In particular, in a hot universe plasma ($T> 10\thinspace\text{MeV}$) the generation of CMF having a maximum magnetic helicity was based on the quantum chiral (Adler) anomaly in relativistic QED plasma for which the difference of right- and left-chiral electron chemical potentials $\Delta \mu=\mu_\mathrm{R} - \mu_\mathrm{L}$ is not equal to zero, $\Delta \mu\neq 0$ \cite{Boyarsky:2011uy}. In the Standard Model (SM) plasma accounting for weak interactions one suggests the magnetic field generation based on the parity violation Chern-Simons (CS) term in the photon self-energy (PSE) \cite{NivSah05,BoyRucSha12}.

Another problem concerns strong magnetic fields existing in neutron stars as remnants of supernovae (SN). In particular, we are interested here how the strongest magnetic fields observed in magnetars~\cite{Duncan} can be generated. To solve this problem, it was recently suggested to use the chiral plasma instability \cite{Yamamoto1,Yamamoto2} caused by an imbalance between right- and left-handed electrons $\mu_\mathrm{R} - \mu_\mathrm{L}=\mu_5\neq 0$ arising via the left-handed electron capture by protons inside the SN core (urca process). Obviously this mechanism is similar to the generation of helical magnetic fields in a hot plasma~\cite{Boyarsky:2011uy}. The chirality flip in both dense media (cases of a hot plasma $T\gg [m_e, \Delta\mu]$ and a degenerate ultrarelativistic electron gas $\mu_5\gg [T, m_e]$)  leads to the damping $\Delta\mu\to 0$, $\mu_5\to 0$ due to collisions that should be taken into account for estimates of the magnetic field generation efficiency.

In the present work we study magnetic field generation problems (both in the early universe and in a supernova) based on the use of the photon PSE in electroweak plasma where a parity violating neutrino interaction with charged leptons is present. Such a contribution to PSE is equivalent to the appearance of the
CS term $\Pi_2(\mathbf{A}\cdot\mathbf{B})$
in the effective Lagrangian of the electromagnetic
field, where $\Pi_2$ is a master parameter
we are looking for to solve a problem of the magnetic field generation, $\mathbf{B}=\nabla\times \mathbf{A}$ is
the magnetic field, and $\mathbf{A}$ is the vector potential.

We shall describe the interaction between neutrinos and charged leptons
in frames of the Fermi theory which is a good approximation at low energies.
Since we study a $\nu\bar{\nu}$ gas embedded into lepton plasma we can treat neutrinos (antineutrinos) as
proper combinations of the external neutrino hydrodynamic currents coming from the effective SM Lagrangian for the $\nu l$ interaction that is linear in the Fermi constant ($\sim G_\mathrm{F}$) being averaged over the neutrino ensemble. Thus, our approach is analogous to the generalized Furry representation in quantum electrodynamics.

Our work is organized as follows. First, in section~\ref{OFFSHELL} we derive the contribution of {\it virtual} charged leptons to the one loop PSE using the exact propagator
of a charged lepton  calculated in Appendix~\ref{PROP} via the effective $\nu l$ interaction in the presence of the neutrino-antineutrino gas. Then, in section~\ref{PLASMA}, using the imaginary time perturbation theory, we calculate the most general plasma contribution to PSE. We analyze our results for the cases of a classical plasma with low temperature and density as well as for hot and degenerate relativistic plasmas.

In the main section~\ref{APPL} we consider some applications of PSE for the evolution of magnetic fields in relativistic plasmas of a supernova and a hot plasma of the early universe (sections~\ref{SUPERNOV} and~\ref{EARLYUNI}).
For such media filled by a plenty of neutrinos we reveal the instability of  $\bf B$-field driven by the neutrino asymmetry $\Delta n_{\nu}= n_{\nu} - n_{\bar{\nu}}\neq 0$.
The analysis of the instabilities of CMF and magnetic fields in a supernova relies on a particular solution of the Faraday equation  modified in SM  that governs the ${\bf B}$-field evolution. Such a solution for a 3D configuration of magnetic field with the maximum magnetic helicity is derived in Appendix~\ref{SYSTEM}. In section~\ref{COMPARISON} we compare our results with some issues in papers based on  chiral properties of ultrarelativistic plasmas.

Finally, in section~\ref{CONCL} we summarize our results and
compare our calculations of the master parameter $\Pi_2$ valid for any plasma with the similar results obtained by other authors.  Some useful formulas of the dimensional regularization are provided in Appendix~\ref{DIMREG} and the example of the calculation of an integral involving plasma effects is given in Appendix~\ref{FTFTINT}.
\section{Photon polarization tensor in a $\nu\bar{\nu} $ gas\label{OFFSHELL}}

In this section we calculate the parity violating term in the polarization tensor in the presence of a $\nu\bar{\nu} $ gas. It should be noted that photons do not interact directly with neutrinos since latter particles are neutral. Thus the $\nu\gamma$ interaction should be mediated by charged leptons, denoted as $l$, which are taken to be virtual particles in this section. We shall take into account the $\nu l$ interaction in propagators of $l$'s as the external mean fields $f^{\mu}_\mathrm{L,R} = (f^0_{\mathrm{L,R}},\mathbf{f}_{\mathrm{L,R}})$ (see Appendix~\ref{PROP}).


We shall be mainly interested in the case of {\it an isotropic} $\nu\bar{\nu}$ gas when $\mathbf{f}_\mathrm{L} = \mathbf{f}_\mathrm{R} = 0$ and the nonzero $f^0_\mathrm{L,R}$ are given in eq.~\eqref{f0LR}. In this situation the most general expression for the polarization tensor reads~\cite{Mohanty:1997mr}
\begin{equation}\label{poltensdec}
  \Pi_{\mu\nu}(k) =
  \left(
    g_{\mu\nu} - \frac{k_\mu k_\nu}{k^2}
  \right)
  \Pi_\mathrm{T} +
  \frac{k_\mu k_\nu}{k^2}
  \Pi_\mathrm{L} +
  \mathrm{i}\varepsilon_{\mu\nu\alpha\beta}k^\alpha (f^\beta_\mathrm{L} - f^\beta_\mathrm{R})
  \Pi_\mathrm{P},
\end{equation}
where $k^\mu = (k^0, \mathbf{k})$ is the photon momentum, $g_{\mu\nu} = \text{diag} (+1, -1, -1, -1)$ is the Minkowski metric tensor, $\varepsilon_{\mu\nu\alpha\beta}$ is the absolute antisymmetric tensor having $\varepsilon^{0123} = +1$, and $\Pi_\mathrm{T,L,P}$ are the form factors of a photon. Since we study parity violating effects, we should analyze the form factor $\Pi_2 = (f^0_\mathrm{L} - f^0_\mathrm{R}) \Pi_\mathrm{P}$. Since only real particles, considered in this section, are neutrinos, we add the superscript ``$\nu$" to photon form factors, e.g., $\Pi_2 \to \Pi_2^{(\nu)}$ etc.

\begin{figure}
  \centering
  \includegraphics[scale=1]{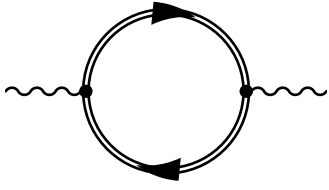}
  \caption{The Feynman diagram for the one loop contribution to PSE given in
  eq.~(\ref{eq:Pivacgen}). The lepton's propagators are shown as broad
  straight lines and correspond to eq.~(\ref{eq:Sexpand}).\label{fig:FeynDiagr}}
\end{figure}

The one loop contribution to PSE is schematically
shown in figure~\ref{fig:FeynDiagr}.
The lepton propagators are represented
as broad lines since we take into account $f^{\mu}_\mathrm{L,R}$ in our calculations.
Note that we shall consider only the contribution to $\Pi_{\mu\nu}^{(\nu)}$ linear in the external fields $f^{\mu}_\mathrm{L,R}$.
The expression for $\Pi_{\mu\nu}^{(\nu)}$, which leads to the nonzero $\Pi_\mathrm{P}^{(\nu)}$ in eq.~\eqref{poltensdec}, reads
\begin{equation}\label{eq:Pivacgen}
  \Pi_{\mu\nu}^{(\nu)} =
  \mathrm{i}e^{2}\int\frac{\mathrm{d}^{4}p}{(2\pi)^{4}}\mathrm{tr}
  \left\{
    \gamma_{\mu}S_{0}(p+k)\gamma_{\nu}S_{1}(p) +
    \gamma_{\nu}S_{0}(p)\gamma_{\mu}S_{1}(p+k)
  \right\},
\end{equation}
where $e$ is the electric charge of $l$ and the propagators $S_{0,1}(p)$ are given in eq.~\eqref{eq:Sexpand}.


The traces of Dirac matrices in eq.~\eqref{eq:Pivacgen} can be evaluated using the following expressions:
\begin{align}\label{traces}
  \mathrm{tr} & (\gamma_{\mu}\gamma_{\alpha}\gamma_{\nu}\gamma_{\beta}\gamma^{5}) =
  -4\mathrm{i}\varepsilon_{\mu\alpha\nu\beta},
  \notag
  \\
  \mathrm{tr} &
  \left\{
    \gamma_{\mu}(\not p+\not k+m)\gamma_{\nu}\sigma_{\alpha\beta}\gamma^{5}(\not p+m)
  \right\}
  \notag
  \\
  & =
  4
  \left\{
    \varepsilon_{\mu\nu\alpha\beta}
    \left[
      m^{2}-p^{2}-(kp)
    \right] -
    k^{\lambda}
    \left[
      \varepsilon_{\alpha\beta\nu\lambda}p_{\mu} -
      \varepsilon_{\alpha\beta\mu\lambda}p_{\nu}
    \right]
  \right\},
  \notag
  \\
  \mathrm{tr} &
  \left\{
    \gamma_{\nu}(\not p-\not k+m)\gamma_{\mu}\sigma_{\alpha\beta}\gamma^{5}(\not p+m)
  \right\}
  \notag
  \\
  & =
  4
  \left\{
    \varepsilon_{\mu\nu\alpha\beta}
    \left[
      p^{2}-m^{2}-(kp)
    \right] -
    k^{\lambda}
    \left[
      \varepsilon_{\alpha\beta\nu\lambda}p_{\mu} -
      \varepsilon_{\alpha\beta\mu\lambda}p_{\nu}
    \right]
  \right\}.
\end{align}
%
To derive eq.~\eqref{traces} we use the fact that $\sigma_{\alpha\beta}\gamma^{5} = \tfrac{\mathrm{i}}{2}\varepsilon_{\alpha\beta\lambda\rho}\sigma^{\lambda\rho}$.

Using the dimensional regularization and eq.~\eqref{intdimreg} we can express $\Pi_2^{(\nu)}$ as
\begin{align}
  \Pi_2^{(\nu)} = &
  (f^0_\mathrm{L} - f^0_\mathrm{R}) \frac{e^{2}}{8\pi^{2}}
  \int_{0}^{1}\mathrm{d}x
  \left(
    \frac{4\pi\lambda^{2}}{M^{2}}
  \right)^{\varepsilon}
  \notag
  \\
  & \times
  \left[
    \Gamma(\varepsilon)(5 - 18x + 12x^2) -
    \Gamma(1+\varepsilon)\frac{8k^{2}x^{2}(1-x)^2}{m^{2}-k^{2}x(1-x)}
  \right],
\end{align}
where $M^{2}=m^{2}-k^{2}x(1-x)$ and $\Gamma(z)$ is the Euler Gamma function. The parameters $\varepsilon$ and $\lambda$ are defined in Appendix~\ref{DIMREG}. Considering the limit $\varepsilon\to0$ and using the fact that $\Gamma(\varepsilon)\approx \tfrac{1}{\varepsilon}+\gamma$, where $\gamma\approx0.577$, we can represent $\Pi_2^{(\nu)}$ in the form,
\begin{equation}\label{eq:Pivacexpl}
  \Pi_2^{(\nu)} = -
  (f^0_\mathrm{L} - f^0_\mathrm{R})
  \frac{e^{2}}{4\pi^{2}}\frac{k^{2}}{m^{2}}
  \int_{0}^{1}\mathrm{d}x\frac{x(1-x)}{1-\frac{k^{2}}{m^{2}}x(1-x)}.
\end{equation}
It should be noted that eq.~\eqref{eq:Pivacexpl} does not contain ultraviolet divergencies.

As shown in ref.~\cite{JacKos99}, the contribution to PSE, calculated in frames of an effective theory which contains a parity violating interaction, is finite but it can depend on the regularization scheme used. Basing on eq.~\eqref{eq:Pivacexpl} we find that $\Pi_2^{(\nu)}=0$ at $k^2=0$, which agrees with the general analysis made in ref.~\cite{ColGla99} for a CPT-odd gauge invariant effective theory. We also note that $\Pi_2^{(\nu)}$ in
eq.~(\ref{eq:Pivacexpl}) coincides with the result of ref.~\cite{Mohanty:1997mr}, where the more fundamental Weinberg-Salam theory was used. Moreover, the fact that $\Pi_2^{(\nu)}$ vanishes at $k^2=0$ also agrees with the finding of ref.~\cite{Gel61},
where it was shown that the neutrino-photon interaction is absent
in the lowest order in the Fermi constant. Nevertheless, as demonstrated
in ref.~\cite{Abbasabadi:2001ps}, the amplitude for $\nu\gamma\to\nu\gamma$
has the nonzero value in two loops.

\section{Plasma contribution to polarization tensor\label{PLASMA}}

In this section we study the direct contribution of charged leptons to the photon form factor $\Pi_2$ corresponding to their parity violating interaction with background neutrinos. We take into account a lepton mass that is absolutely necessary, e.g., for the classical nonrelativistic plasma. Thus leptons are not chirally polarized. For relativistic plasmas we again substitute an effective lepton mass $m_\mathrm{eff}(T,\mu)$~\cite{Braaten:1993jw} in photon dispersion characteristics, see below in eq.~\eqref{meff}, that also differs our approach from the use of the lepton chirality.

Thus, in this section we obtain the general expression for $\Pi_{2}$ taking into account both the temperature and the chemical potential of the charged leptons. It means that these leptons now are not virtual particles. We also exactly account for the photon's dispersion relation $k_0 = k_0(\mathbf{k})$ in this plasma. On the basis of the general results we discuss the cases of low temperature and low density classical plasma, as well as hot relativistic and degenerate relativistic plasmas.

If we study the photon propagation in a plasma of charged leptons with nonzero temperature and density, the photon's dispersion relation differs from the vacuum one, $k^{2}=(k_0^2 - {\bf k}^2)\neq0$. As seen in eq.~(\ref{eq:Pivacexpl}), in this case $\Pi_2^{(\nu)} \neq 0$. However, we should also evaluate the direct contribution of plasma particles to the parity violating form factor of a photon. We can define it as $\Pi_{2}^{(\nu l)}$ analogously to section~\ref{OFFSHELL}. Therefore we shall study the system consisting of a real $l$'s plasma and a real $\nu\bar{\nu}$ gas. The presence of $\nu$'s and $\bar{\nu}$'s is essential since it is these particles which provide the nonzero contribution to the parity violating form factor based on the $\nu l$ interaction.

The expression for the contribution to PSE from the plasma
of not virtual leptons can be obtained if we make the following replacement
in eq.~(\ref{eq:Pivacgen}) (see ref.~\cite{KapGal06}):
\begin{equation}\label{Mfintr}
  \mathrm{i}\int\frac{\mathrm{d}p^{0}}{2\pi}\to T\sum_{n},
  \quad
  p^{0}=(2n+1)\pi T\mathrm{i}+\mu,
  \quad
  n=0,\pm1,\pm2,\dotsc,
\end{equation}
where $T$ and $\mu$ are the temperature and the chemical potential
of the $l$'s plasma. In principle, we can discuss a general situation when $T$ and $\mu$ are different from $T_{\nu_\alpha}$ and $\mu_{\nu_\alpha}$ defined in eq.~\eqref{nnugen}. However, in section~\ref{APPL}, where we study the application of our calculations, the system in the thermodynamic equilibrium is considered. Thus, in the following we shall suppose that $T = T_\nu$, where $T_\nu$ is the $\nu\bar{\nu}$ gas temperature equal for all neutrino flavors. However, we shall keep different $\mu$ and $\mu_{\nu_\alpha}$.


Using eqs.~\eqref{eq:Pivacgen}, \eqref{traces}, \eqref{Mfintr},  and~\eqref{eq:Sexpand},
as well as defining the effective chemical potentials $\mu^{\pm} = \mu - (f^0_\mathrm{L} + f^0_\mathrm{R})/2 \pm k_{0}x$,
we can express $\Pi_{2}^{(\nu l)}$ in the following form:
\begin{align}\label{Pi2Tgen}
  \Pi_{2}^{(\nu l)} = &
  - \frac{e^{2}(f^0_\mathrm{L} - f^0_\mathrm{R})}{2}
  \int_{0}^{1}dx
  \int\frac{\mathrm{d}^{3}p}{(2\pi)^{3}}
  \frac{1}{\mathcal{E}_{\mathbf{p}}^{3}}
  \notag
  \\
  & \times
  \bigg\{
    \frac{1}{\exp[\beta(\mathcal{E}_{\mathbf{p}}-\mu^{+})]+1} +
    \frac{1}{\exp[\beta(\mathcal{E}_{\mathbf{p}}+\mu^{+})]+1}
    \notag
    \\
    & +
    \frac{\beta\mathcal{E}_{\mathbf{p}}}{2}
    \left[
      \frac{1}{\cosh[\beta(\mathcal{E}_{\mathbf{p}}-\mu^{+})]+1} +
      \frac{1}{\cosh[\beta(\mathcal{E}_{\mathbf{p}}+\mu^{+})]+1}
    \right]
    \notag
    \\
    & -
    (1-x)
    \bigg[
    \frac{1}{\mathcal{E}_{\mathbf{p}}^2}
    \left(
      \mathbf{p}^{2}
      \left[
        1-\frac{5}{3}x
      \right] -
      \left[
        k^{2}x(1-x)+m^{2}
      \right]x
    \right)
    \left(
      J_{0}^{(+)}+J_{0}^{(-)}
    \right)
    \notag
    \\
    & +
    \beta k^{0} x(1-2x)
    \left(
      J_{1}^{(+)}-J_{1}^{(-)}
    \right) +
    x
    \left(
      J_{2}^{(+)}+J_{2}^{(-)}
    \right)
  \bigg]
  \bigg\},
\end{align}
where
\begin{align}\label{J012}
  J_{0}^{(\pm)} = &
  3
  \bigg\{
    \frac{1}{\exp[\beta(\mathcal{E}_{\mathbf{p}}+\mu^{\pm})]+1} +
    \frac{1}{\exp[\beta(\mathcal{E}_{\mathbf{p}}-\mu^{\pm})]+1}
    \notag
    \\
    & +
    \frac{\beta\mathcal{E}_{\mathbf{p}}}{2}
    \left[
      \frac{1+\beta\mathcal{E}_{\mathbf{p}}
      \tanh[\beta(\mathcal{E}_{\mathbf{p}} + \mu^{\pm})/2]/3}
      {1+\cosh[\beta(\mathcal{E}_{\mathbf{p}}+\mu^{\pm})]} +
      \frac{1+\beta\mathcal{E}_{\mathbf{p}}
      \tanh[\beta(\mathcal{E}_{\mathbf{p}}-\mu^{\pm})/2]/3}
      {1+\cosh[\beta(\mathcal{E}_{\mathbf{p}}-\mu^{\pm})]}
    \right]
  \bigg\},
  \notag
  \\
  J_{1}^{(\pm)} = & -\frac{1}{2}
  \left\{
    \frac{1+\beta\mathcal{E}_{\mathbf{p}}
    \tanh[\beta(\mathcal{E}_{\mathbf{p}}+\mu^{\pm})/2]}
    {1+\cosh[\beta(\mathcal{E}_{\mathbf{p}}+\mu^{\pm})]} -
    \frac{1+\beta\mathcal{E}_{\mathbf{p}}
    \tanh[\beta(\mathcal{E}_{\mathbf{p}}-\mu^{\pm})/2]}
    {1+\cosh[\beta(\mathcal{E}_{\mathbf{p}}-\mu^{\pm})]}
  \right\},
  \notag
  \\
  J_{2}^{(\pm)} = &
  -
  \bigg\{
    \frac{1}{\exp[\beta(\mathcal{E}_{\mathbf{p}}+\mu^{\pm})]+1} +
    \frac{1}{\exp[\beta(\mathcal{E}_{\mathbf{p}}-\mu^{\pm})]+1}
    \notag
    \\
    & +
    \frac{\beta\mathcal{E}_{\mathbf{p}}}{2}
    \left[
      \frac{1-\beta\mathcal{E}_{\mathbf{p}}
      \tanh[\beta(\mathcal{E}_{\mathbf{p}}+\mu^{\pm})/2]}
      {1+\cosh[\beta(\mathcal{E}_{\mathbf{p}}+\mu^{\pm})]} +
      \frac{1-\beta\mathcal{E}_{\mathbf{p}}
      \tanh[\beta(\mathcal{E}_{\mathbf{p}}-\mu^{\pm})/2]}
      {1+\cosh[\beta(\mathcal{E}_{\mathbf{p}}-\mu^{\pm})]}
    \right]
  \bigg\}.
\end{align}
Here $\mathcal{E}_{\mathbf{p}} = \sqrt{\mathbf{p}^2 + M^2}$, $\beta = 1/T$, and $M^2$ is defined in section~\ref{OFFSHELL}. To obtain eqs.~\eqref{Pi2Tgen} and~\eqref{J012} we assume that $k^2 < 4 m^2$, i.e. no creation of $l\bar{l}$-pairs occurs.

It should be noted that in deriving eqs.~\eqref{Pi2Tgen} and~\eqref{J012} we exactly account for the $l$'s mass $m$. Thus charged leptons are not taken to be chirally polarized. It means that the magnetic field instability discussed later in section~\ref{APPL}, which results from
the nonzero $\Pi_{2} = \Pi_{2}^{(\nu)} + \Pi_{2}^{(\nu l)} $, is generated rather by the neutrino asymmetry $(n_{\nu} - n_{\bar{\nu}})\neq 0$ than by the chiral asymmetry of charged leptons $\sim (\mu_\mathrm{R} - \mu_\mathrm{L})\neq 0$ studied, e.g., in refs.~\cite{Boyarsky:2011uy,Yamamoto1,Yamamoto2}.

\subsection{Low density classical plasma\label{CLASSPLASM}}

Let us first discuss the case of a low density plasma
of $l$'s, that corresponds to $k^{2}\ll m^{2}$.
Using the general eqs.~\eqref{Pi2Tgen} and~\eqref{J012} in the limit $\max(k_0^{2},\mathbf{k}^2) \ll m^{2}$ we obtain that
\begin{align}\label{eq:Pi2T}
  \Pi_{2}^{(\nu l)}=
  &
  -\frac{7}{6} e^{2} (f^0_\mathrm{L} - f^0_\mathrm{R})
  \int\frac{\mathrm{d}^{3}p}{(2\pi)^{3}}\frac{1}{\mathcal{E}_{\mathbf{p}}^{3}}
  \nonumber
  \\
  &
  \times
  \bigg\{
    \frac{m^{2}}{\mathcal{E}_{\mathbf{p}}^{2}}
    \left[
      \frac{1}{\exp[\beta(\mathcal{E}_{\mathbf{p}}-\mu)]+1} +
      \frac{1}{\exp[\beta(\mathcal{E}_{\mathbf{p}}+\mu)]+1}
    \right]
    \nonumber
    \\
    & +
    \frac{m^{2}\beta}{2\mathcal{E}_{\mathbf{p}}}
    \left[
      \frac{1}{\cosh[\beta(\mathcal{E}_{\mathbf{p}}-\mu)]+1} +
      \frac{1}{\cosh[\beta(\mathcal{E}_{\mathbf{p}}+\mu)]+1}
    \right]
    \nonumber
    \\
    & -
    \frac{\beta^{2}\mathbf{p}^{2}}{6}
    \left[
      \frac{\tanh[\beta(\mathcal{E}_{\mathbf{p}}-\mu)/2]}
      {\cosh[\beta(\mathcal{E}_{\mathbf{p}}-\mu)]+1} +
      \frac{\tanh[\beta(\mathcal{E}_{\mathbf{p}}+\mu)/2]}
      {\cosh[\beta(\mathcal{E}_{\mathbf{p}}+\mu)]+1}
    \right]
  \bigg\} ,
\end{align}
where $\mathcal{E}_{\mathbf{p}} = \sqrt{\mathbf{p}^2 + m^2}$ since we neglect the photon's dispersion in plasma. Note that $\Pi_{2}^{(\nu l)}$ in eq.~\eqref{eq:Pi2T} exactly accounts for $T$ and $\mu$.

To estimate the values of $\Pi_2^{(\nu)}$ and $\Pi_{2}^{(\nu l)}$,
we shall consider the low temperature limit: $T\ll m$. We will identify
$l$ with an electron and assume that the electron gas has a classical
Maxwell distribution. For this medium we get that $k^{2}=4\pi\alpha_{\mathrm{em}}n_{e}/m$,
where $\alpha_{\mathrm{em}}=e^{2}/4\pi = 1/137$ is the fine structure constant
and $n_{e}$ is the background electron density. Moreover for a classical
electron gas one has that $\mu=m+T\ln\left[\tfrac{n_{e}}{g_{s}}\left(\tfrac{2\pi}{mT}\right)^{3/2}\right]$,
where $g_{s}=2$ is the number of spin degrees of freedom of an electron.
Using eqs.~(\ref{eq:Pivacexpl}) and~(\ref{eq:Pi2T}), we get that
\begin{equation}\label{eq:lowtempclass}
  \tilde{\Pi}_2^{(\nu)} =
  -\frac{2\alpha_{\mathrm{em}}^{2}}{3}
  (f^0_\mathrm{L} - f^0_\mathrm{R})
  \frac{n_{e}}{m^{3}},
  \quad
  \tilde{\Pi}_{2}^{(\nu l)} =
  -\frac{7\pi\alpha_{\mathrm{em}}}{3}
  (f^0_\mathrm{L} - f^0_\mathrm{R})
  \frac{n_{e}}{m^{3}},
\end{equation}
where we add a tilde over $\Pi_2^{(\nu,\nu l)}$ to stress that these quantities correspond to real photons in plasma (plasmons) rather to the virtual photons. In the following we shall omit the tilde in order not to encumber notations.

One can see that $\Pi_{2}^{(\nu l)}$ in eq.~(\ref{eq:lowtempclass})
is $\tfrac{7\pi}{2\alpha_{\mathrm{em}}}\sim10^{3}$ times greater than
$\Pi_2^{(\nu)}$. Note that for a classical nonrelativistic plasma, corresponding to $m \gg \max(|\mathbf{p}|, T)$, the integrals in last two lines in eq.~\eqref{eq:Pi2T} cancel each other while the integral in the first line leads to the term $\Pi_2^{(\nu l)}$ in eq.~\eqref{eq:lowtempclass}.

Let us study the derived $\Pi_{2}$ in the static limit $k_0 = 0$. If we discuss the situation when only virtual charged leptons contribute to PSE, we should set $n_e \to 0$ in eq.~\eqref{eq:lowtempclass}. This limit is equivalent to $k^2 \to 0$. Using eq.~\eqref{eq:lowtempclass}, we obtain that $\Pi_2 \to 0$. This our result is in agreement the findings
of ref.~\cite{Gel61}, where it was found that the one loop contribution to $\nu\gamma$-interaction should be vanishing. The leading nonzero contribution to $\Pi_2$ in the case when charged leptons are virtual particles, i.e. when we neglect the plasma contribution given by eqs.~\eqref{Pi2Tgen} and~\eqref{J012}, was obtained in ref.~\cite{Abbasabadi:2001ps}. Using the results of ref.~\cite{Abbasabadi:2001ps}, one gets that in this situation $\Pi_{2} \sim \alpha_{\mathrm{em}} G_{\mathrm{F}} / M_{\mathrm{W}}^{4}$, where $M_{\mathrm{W}}$ is the $W$-boson mass.

\subsection{Hot relativistic plasma\label{HOTRELPLAS}}

The dispersion relation for transverse waves in relativistic plasma reads~\cite{Braaten:1993jw},
\begin{equation}\label{relpldispgen}
  k_0^2 = \mathbf{k}^2 + \omega_p^2
 \left( \frac{3k_0^2}{2\mathbf{k}^2}\right)
  \left[
    1-\frac{(k_0^{2}-\mathbf{k}^{2})}{k_0^{2}}\left(
    \frac{k_0}{2|\mathbf{k}|}\right)
    \ln\frac{k_0+|\mathbf{k}|}{k_0-|\mathbf{k}|}
  \right].
\end{equation}
The plasma frequency $\omega_p$ can be found from the following expression:
\begin{equation}\label{omegap}
  \omega_{p}^{2} =
  \frac{4\alpha_{\mathrm{em}}}{\pi}
  \int_{0}^{\infty}\mathrm{d}p
  \frac{\mathbf{p}^2}{\mathcal{E}_{\mathbf{p}}^2}
  \left(
    1-\frac{v^{2}}{3}
  \right)
  \left[
    \frac{1}{\exp[\beta(\mathcal{E}_{\mathbf{p}}+\mu)]+1} +
    \frac{1}{\exp[\beta(\mathcal{E}_{\mathbf{p}}-\mu)]+1}
  \right],
\end{equation}
where $v = |\mathbf{p}|/\mathcal{E}_{\mathbf{p}}$.

Using eq.~\eqref{omegap} in the relativistic limit $T \gg \max(m,\mu)$, we get that
$\omega_p^2 = 4 \alpha_{\mathrm{em}} \pi T^2/9$. The transcendant eq.~\eqref{relpldispgen} can be explicitly solved if long waves with $k_0^2 \gg \mathbf{k}^2$ are considered. In this situation the dispersion relation is $k^2 = \omega_p^2$.

It should be noted that the electron's mass in plasma can significantly differ from its vacuum value. The radiative corrections to the electron's mass were studied in ref.~\cite{Kli82}. Thus, if we consider a dense and hot plasma, we should replace \footnote{Under intermediate conditions $m_e\sim m_\mathrm{eff}$ in plasma with $\mu\neq 0$ or $T\neq 0$ (or both) the effective mass of an electron should be $m_e/2 + (m_e^2/4 + m_\mathrm{eff}^2)^{1/2}$, see in ref.~\cite{Braaten:1993jw}.}
\begin{equation}\label{meff}
  m^2 \to m_\mathrm{eff}^2 = \frac{e^2}{8\pi^2}
  (\mu^2 + \pi^2 T^2),
\end{equation}
in eqs.~\eqref{eq:Pivacexpl}, \eqref{Pi2Tgen}, and~\eqref{J012}. Note that eq.~\eqref{meff} is valid for both $T \gg \mu$ and $\mu \gg T$. Accounting for the dispersion relation and the expression for $\omega_p$, we get that $k^2 < 4 m_\mathrm{eff}^2$ in a hot relativistic plasma.

\begin{figure}
  \centering
  \includegraphics[scale=1]{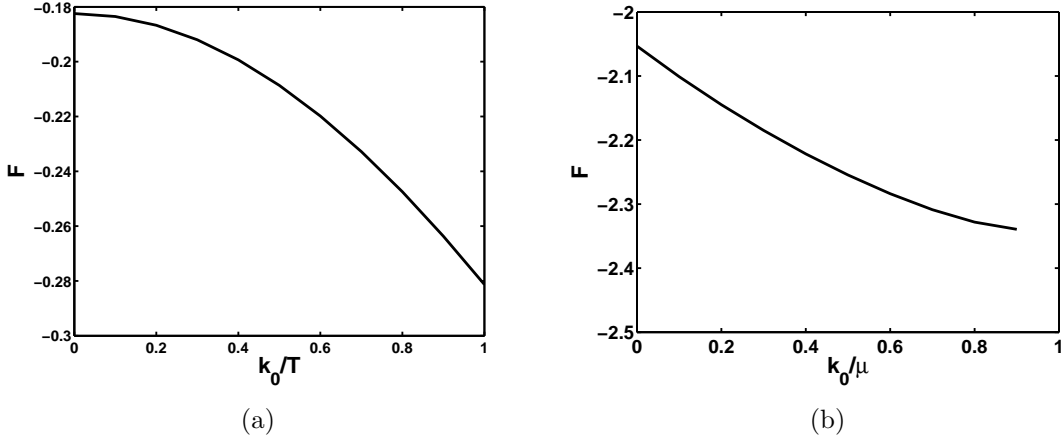}
  \caption{The function $F$ versus $k_0$. (a) Hot relativistic plasma. (b) Degenerate relativistic plasma.\label{FTmu}}
\end{figure}

Let us represent $\Pi_2$ as
\begin{equation}\label{Fdef}
  \Pi_{2}=\frac{\alpha_{\mathrm{em}}}{\pi}(f^0_\mathrm{L} - f^0_\mathrm{R})F,
\end{equation}
where $F$ is the dimensionless function which depends on $k_0/T$. Note that $\Pi_{2}$ in eq.~\eqref{Fdef} includes the contributions from eqs.~\eqref{eq:Pivacexpl} and~\eqref{Pi2Tgen}. Accounting for eq.~\eqref{meff}, we present the behaviour of $F$ versus we $k_0/T$ in figure~\ref{FTmu}(a). We study long waves limit when $k_0 \approx \omega_p \approx 0.1 T$. Thus we should be interested in the values of $F$ corresponding to $k_0 \ll T$.
%

One can see in figure~\ref{FTmu}(a) that for a hot relativistic plasma $\Pi_2$ is nonvanishing in the static limit: $F(k_0 \to 0) \approx -0.18$. However this nonzero value strongly depends on the photon's dispersion law in such a plasma.

\subsection{Degenerate relativistic plasma\label{DEGRELPLAS}}

In case of a degenerate plasma the dispersion relation for transverse waves is~\cite{Braaten:1993jw},
\begin{equation}\label{degpldispgen}
  k_0^2 = \mathbf{k}^2 + \omega_p^2
 \left( \frac{3k_0^2}{2v_\mathrm{F}^2\mathbf{k}^2}\right)
  \left[
    1-\frac{(k_0^{2}-v_\mathrm{F}^2\mathbf{k}^{2})}{k_0^{2}}
    \left(\frac{k_0}{2v_\mathrm{F}|\mathbf{k}|}\right)
    \ln\frac{k_0+ v_\mathrm{F}|\mathbf{k}| }{k_0-v_\mathrm{F}|\mathbf{k}| }
  \right],
\end{equation}
where $v_\mathrm{F}$ is the Fermi velocity. The plasma frequency $\omega_p$ can be found from eq.~\eqref{omegap} if we make the following replacement: $\{\exp[\beta(\mathcal{E}_{\mathbf{p}}-\mu)]+1\}^{-1} \to \theta(\beta[\mu - \mathcal{E}_{\mathbf{p}}])$ and $\{\exp[\beta(\mathcal{E}_{\mathbf{p}}+\mu)]+1\}^{-1} \to 0$, where $\theta(z)$ is the Heaviside step function.

Let us discuss the degenerate plasma in the relativistic limit. In this situation $v_\mathrm{F} = 1$ and $\omega_p^2 = 4 \alpha_{\mathrm{em}} \mu^2 / 3 \pi$. The general dispersion relation in eq.~\eqref{degpldispgen} transforms into $k^2 = \omega_p^2$ if we study long waves. Using eq.~\eqref{meff}, we also get the effective electron mass in a degenerate plasma. One can check that the inequality $k^2 < 4 m_\mathrm{eff}^2$ is valid.

Using eqs.~\eqref{eq:Pivacexpl}, \eqref{Pi2Tgen}, and~\eqref{J012} in the limit $\mu \gg T$, as well as the following representations of the Dirac delta function and its derivative:
\begin{equation}
  \lim_{\beta\to\infty} \frac{\beta}{\cosh(\beta x)+1} =
  2\delta(x),
  \quad
  \lim_{\beta\to\infty} \frac{\beta^2 \tanh(\beta x/2)}{\cosh(\beta x)+1} =
  -2\delta'(x),
\end{equation}
we can derive the expression for the function $F = F(k_0/\mu)$, see eq.~\eqref{Fdef}, in case of a relativistic degenerate plasma. In this situation the integration over momenta can be performed explicitly. However, here we do not give the expression for $F$ since it is very cumbersome.


The function $F$ versus $k_0/\mu$ is shown in figure~\ref{FTmu}(b). We discuss the long waves limit. Thus $k_0 \approx \omega_p \approx 0.06 \mu$. It means that for our purposes we should consider $F$ at $k_0 \ll \mu$. One can see in figure~\ref{FTmu}(b) that $F(k_0 \to 0) \approx -2.05$. Therefore, as in case of a hot relativistic plasma, for a degenerate relativistic plasma $\Pi_2$ is nonvanishing at $k_0 \to 0$, but its actual value $\Pi_2(0)$ is different from that found in section~\ref{HOTRELPLAS}.

\section{Instability of magnetic fields in relativistic plasmas driven by neutrino asymmetries \label{APPL}}

We consider below two cases for which the CS term $\Pi_2$ in the photon polarization operator $\Pi_{\mu\nu}$ plays a crucial role. A nonzero $\Pi_2$ leads to the $\alpha$-dynamo amplification (instability) of a seed magnetic field even without fluid vortices or any rotation $\Omega$ in plasma which are usually exploited in the standard MHD approach for $\alpha \Omega$-dynamo \cite{1983flma....3.....Z}. The first case considered here concerns the magnetic field growth in a degenerate ultrarelativistic electron plasma, $\mu\gg \max(T, m_e)$, during the collapse and deleptonization phases of a supernova burst.  In the second case we consider below a hot plasma of the early universe with the temperatures $T\gg \max(m_e, \mu)$ before the neutrino decoupling at $T>T_\mathrm{dec}\simeq 2\div 3\thinspace\text{MeV}$. In both cases neutrinos are in equilibrium with a plasma environment. For these applications we use our result in eq.~(\ref{Fdef}).

First, we derive in subsection \ref{Faradayeq} the Faraday equation generalized in SM to find the key parameters leading to the ${\bf B}$-field instability. The corresponding evolution equations for the spectra of the magnetic helicity density $h(k,t)$ and magnetic energy density $\rho_\mathrm{B}(k,t)$ presented in Appendix \ref{SYSTEM} allow us to interpret the simplest solution of Faraday equation for the case of the maximum helicity density obeying the inequality $h(k,t)\leq 2\rho_\mathrm{B}(k,t)/k$ \cite{Biskamp}. Here $h(t)=\int \mathrm{d}k h(k,t)=V^{-1}\int \mathrm{d}^3 x ({\bf A}\cdot{\bf B})$ is the magnetic helicity density and
 $\rho_\mathrm{B}(t)=\int \mathrm{d}k\rho_\mathrm{B}(k,t)=B^2(t)/2$ is the magnetic energy density for an uniform isotropic medium.

An excess of electron neutrinos during
a first second of a supernova explosion\footnote{Neutrino emission prevails over the antineutrino one during first milliseconds of a supernova burst due to the reaction $e^- + p\to n + \nu_e$ (urca-process) before its equilibrium with beta-decays $n\to p + e^- + \bar{\nu}_e$ is settled in (see figure~11.3 in ref.~\cite{Raffelt}).} allow us to put $n_{\nu_e} - n_{\bar{\nu}_e}\neq 0$ in the problem of the magnetic field amplification considered in subsection~\ref{SUPERNOV}.
In subsection \ref{EARLYUNI} we find the lower bound on the neutrino asymmetry providing the growth of CMF field in our causal scenario. It would be interesting to compare such a limit with the upper bound on the electron neutrino-antineutrino asymmetry $| \xi_{\nu_e}| \leq 0.07$ given by the Big Bang nucleosynthesis (BBN) constraint~\cite{Dolgov:2002ab}.
Thus, we shall consider magnetic fields in media with a plenty of neutrinos (antineutrinos) where a nonzero neutrino asymmetry exists. Finally, in section~\ref{COMPARISON} we compare our findings with what other authors found in similar problems.

\subsection{Generalized Faraday equation in the Standard Model \label{Faradayeq}}

The existence of a neutrino asymmetry accounting for the difference in eq.~(\ref{f0LR}),
\begin{equation}\label{deltaf}
f_\mathrm{L}^0 - f_\mathrm{R}^0=2\sqrt{2}G_\mathrm{F}\left[\Delta n_{\nu_e} - \frac{1}{2}\sum_{\alpha}\Delta n_{\nu_{\alpha}}\right],
\end{equation}
leads to a non-zero parity violation term in the photon polarization operator $\Pi_{ij}(\omega, {\bf k})=\mathrm{i}\varepsilon_{ijn}k^n\Pi_2(\omega, k)$, where $\Pi_2$ is given by eq.~(\ref{Fdef}) and $\omega \equiv k_0$.

The CS polarization term in eq.~(\ref{Fdef}) corresponds to the induced pseudovector current in the Fourier representation,
\begin{equation}
  {\bf j}_5(\omega, {\bf k})=\Pi_2(\omega, k){\bf B}(\omega, {\bf k}),
\end{equation}
entering the generalized Maxwell equation in the standard model (SM)
\begin{equation}\label{Maxwell}
  \mathrm{i}{\bf k}\times {\bf B}(\omega, {\bf k}) +
  \mathrm{i}\omega{\bf E}(\omega, {\bf k}) =
  {\bf j}(\omega, {\bf k}) + {\bf j}_5(\omega, {\bf k}).
\end{equation}
Expressing the ohmic current as ${\bf j}(\omega, {\bf k})=\sigma_\mathrm{cond}{\bf E}(\omega, {\bf k})$,
then neglecting the displacement current in the l.h.s. of eq.~(\ref{Maxwell}), that is a standard assumption in the MHD approach for which $\omega\ll \sigma_\mathrm{cond}$~\footnote{The conductivity $\sigma_\mathrm{cond}=\omega_p^2/\nu_\mathrm{coll} = 4\pi\alpha_\mathrm{em}T^2/9\nu_\mathrm{coll}\sim T/\alpha_\mathrm{em}\sim 100T$ depends on the Coulomb collision frequency $\nu_\mathrm{coll}=\sigma_\mathrm{Coul}n_e=[4\pi L\alpha_\mathrm{em}^2/9T^2]n_e\sim \alpha_\mathrm{em}^2T$. Here we use the values for the electron density $n_e=0.183T^3$ in a hot plasma and $L\sim 10$ for the Coulomb logarithm. Obviously the MHD condition $\omega=\omega_t\ll \sigma_\mathrm{cond}$ is fulfilled to obtain eq.~(\ref{Faraday}).}, and finally using the Bianchi identity ${\bf k}\times {\bf E}=\omega {\bf B}$, one gets the generalized Faraday equation in SM in the coordinate representation,
\begin{equation}\label{Faraday}
\frac{\partial {\bf B}}{\partial t}= \alpha\nabla\times {\bf B} +  \eta\nabla^2{\bf B},
\end{equation}
where $\alpha$ is the magnetic helicity parameter,
\begin{equation}\label{helicity}
\alpha=\left(\frac{\Pi_2}{\sigma_\mathrm{cond}}\right),
\end{equation}
and $\eta=(\sigma_\mathrm{cond})^{-1}$ is the magnetic diffusion coefficent.

Here we use the long-wave approximation for large-scale magnetic fields where the operator $\Pi_2(k_0, k= 0)$ is at least uniform, $k\to 0$, and almost stationary since the function $F(x)$ depends on a small ratio $x=k_0/T\ll 1$ or $x=k_0/\mu\ll 1$. For instance, in the long-wave limit $k\ll \omega_t$ the transversal plasmons (photons) have the spectrum $k_0^2\equiv\omega_t^2=\omega_p^2 + \mathbf{k}^2\approx \omega_p^2=4\pi\alpha_\mathrm{em}T^2/9$ in a hot plasma ($T\gg \max[\mu, m_e$]) and  $k_0^2\equiv\omega_t^2=\omega_p^2 + \mathbf{k}^2\approx \omega_p^2=4\alpha_\mathrm{em}\mu^2/3\pi$ in the ultrarelativistic degenerate electron gas ($\mu\gg \max[m_e,T]$) \cite{Braaten:1993jw} (see spectra in eqs.~(\ref{relpldispgen}) and~(\ref{degpldispgen}) above). In a relativistic plasma this approximation corresponds to the negligible spatial dispersion, $k_0\gg k\langle v\rangle\sim k$, where we put $v=1$ both in hot and degenerate relativistic plasmas. Here $k=|\mathbf{k}|$ is the wave number. Thus, the ratio $k_0/T\sim 0.1$ or $k_0/\mu\sim 0.06$ allows us to consider $\Pi_2\approx \text{const}$ without temporal and spatial dispersion as a function of the temperature $T$ (a hot plasma in the early universe) or the chemical potential $\mu$ (a degenerate electron gas in a supernova) only.

\subsection{Amplification of a seed magnetic field  in a supernova\label{SUPERNOV}}

During the collapse (time $t<0.1\thinspace\text{s}$ after onset of collapse, see figure~11.1 in ref.~\cite{Raffelt}) one can neglect $\nu_{\mu,\tau}$ emission and $\Pi_2$ reads
\begin{equation}\label{Pi2}
\Pi_2(k_0, 0)=\left[\frac{\sqrt{2}\alpha_\mathrm{em}G_\mathrm{F}(n_{\nu_e} - n_{\bar{\nu}_e})}{\pi}\right] F\left(k_0/\mu\right),
\end{equation}
where the function $F(x)$ is shown in figure~\ref{FTmu}(b)
for a degenerate ultrarelativistic electron gas with $\mu\gg \max(T,m_e)$.

Let us give some estimates for $\Pi_2$ in a collapsing SN with the progenitor stellar mass $M\sim 8M_{\odot}$ considered in ref.~\cite{Raffelt} (see there the plots for evolution stages in figures~11.1-11.3). In order to obtain $\Pi_2$ we should find the appropriate neutrino asymmetry density $\Delta n_{\nu_e}$.

At the stage just after collapse neutrinos are captured, their free path does not exceed the core radius $\lambda_{\nu}\ll R\sim 10\thinspace\text{km}$. For instance, for the nuclear core density $\rho_0=M_\mathrm{N}(n_n + n_p)=3\times 10^{14}\thinspace\text{g} \cdot \text{cm}^{-3}$ one gets $\lambda_{\nu}\sim 300\thinspace\text{m}$ if $E_{\nu}=30\thinspace\text{MeV}$, or $\lambda_{\nu}\sim 2.7\thinspace\text{km}$ for $E_{\nu}=10\thinspace\text{MeV}$. Here, using the nucleon mass $M_\mathrm{N}=940\thinspace\text{MeV}$, one gets the baryon density $n_\mathrm{B}=n_p + n_p=1.8\times 10^{38}\thinspace\text{cm}^{-3}$.

The lepton abundance $Y_\mathrm{L}=0.3$ is typical for the material in a SN core just after the collapse, so that the equilibrium condition $Y_\mathrm{L} n_\mathrm{B} = n_e + n_{\nu_e}=[p_{\mathrm{F}_e}^3 + p_{\mathrm{F}_{\nu_e}}^3]/3\pi^2=5.4\times 10^{37}\thinspace\text{cm}^{-3}$ allows us to look for the Fermi momenta for degenerate electrons and neutrinos, $p_{\mathrm{F}_e}$ and $p_{\mathrm{F}_{\nu_e}}$. The second equation for these quantities comes from the consideration of figure~D7(a) in ref.~\cite{Raffelt}, where for the same matter density $\rho_0=3\times 10^{14}\thinspace\text{g}\cdot \text{cm}^{-3}$ one finds the difference $\mu_n-\mu_p=\mu_e - \mu_{\nu_e}=40\thinspace\text{MeV}$ corresponding to the temperature $T\simeq 10\thinspace\text{MeV}$.
Note that leptons are ultrarelativistic, $\mu_e - \mu_{\nu_e}=p_{\mathrm{F}_e} - p_{\mathrm{F}_{\nu_e}}=40\thinspace\text{MeV}$, while nucleons are degenerate and nonrelativistic, $\mu_n - \mu_p=E_{\mathrm{F}_n} - E_{\mathrm{F}_p}=[p_{\mathrm{F}_n}^2 - p_{\mathrm{F}_p}^2]/2M_\mathrm{N}=40\thinspace\text{MeV}$. Eventually we get all Fermi momenta in such dense core: $p_{\mathrm{F}_{\nu_e}}=163\thinspace\text{MeV}$, $p_{\mathrm{F}_n}=341\thinspace\text{MeV}$, $p_{\mathrm{F}_e}=p_{\mathrm{F}_p}=203\thinspace\text{MeV}$. Here the last equality comes from the electroneutrality condition $n_e=n_p$. Thus, we get
the electron neutrino density at this stage of the SN evolution, $n_{\nu_e}=1.46\times 10^5\thinspace\text{MeV}^3= 1.9\times 10^{37}\thinspace\text{cm}^{-3}$, which should be substituted into eq.~(\ref{Pi2}) neglecting antineutrino contribution.

The magnetic diffusion time  $t_\mathrm{diff}=\Lambda^2/\eta$ seen from the Faraday eq.~(\ref{Faraday}),
\begin{equation}\label{diff}
  t_\mathrm{diff} =
  \frac{\sigma_\mathrm{cond}}{k^2}=\frac{\sigma_\mathrm{cond}}{\Pi_2^2},
\end{equation}
is given by the electrical conductivity for degenerate
ultrarelativistic electrons and degenerate nonrelativistic protons, $\sigma_\mathrm{cond}=\omega_p^2/\nu_\mathrm{coll}$ \cite{Kelly}.
Note that the combined effects of the degeneracy and the shielding reduce the collision frequency $\nu_\mathrm{coll}\sim T^2$. Thus collisions
of charged particles are blocked due to the Pauli principle since states $p< p_\mathrm{F}$ are busy and $\nu_\mathrm{coll}\to 0$ at $T\to 0$.

The electrical conductivity was found in ref.~\cite{Kelly},
\begin{equation}\label{conductivity}
  \sigma_\mathrm{cond} =
  \frac{1.6\times 10^{28}}{(T/10^8\thinspace\mathrm{K})^2}
  \left(
    \frac{n_e}{10^{36}\thinspace\mathrm{cm}^{-3}}
  \right)^{3/2}\thinspace\mathrm{s}^{-1}.
\end{equation}
For $p_{\mathrm{F}_e}=203\thinspace\text{MeV}$ and the corresponding electron density $n_e=p_{\mathrm{F}_e}^3/3\pi^2=3.7\times 10^{37}\thinspace\text{cm}^{-3}$, as well as the temperature $T=10\thinspace\text{MeV}\simeq 10^{11}\thinspace\text{K}$ in SN core we have just estimated,  eq.~\eqref{conductivity} gives $\sigma_\mathrm{cond}=2250\thinspace\text{MeV}$. This result leads to the estimate $t_\mathrm{diff}=0.023\thinspace\text{s}$. It means that any seed magnetic field $B_0$ existing in plasma does not dissipate ohmically during first milliseconds after onset of collapse, $t\ll t_\mathrm{diff}$, and evolves for a given wave number $k$ through the $\alpha$-dynamo driven by neutrino asymmetries (see in Appendix \ref{SYSTEM})
as
\begin{equation}\label{Balpdyn}
  B(t,k)=B_0\exp
  \left[
    \int_{t_0}^t(|\alpha| k - \eta k^2)\mathrm{d}t'
  \right].
\end{equation}
If $k<|\alpha|/\eta = |\Pi_2|$, the seed magnetic field in eq.~\eqref{Balpdyn} will grow exponentially. The fastest growth corresponds to the $\alpha^2$-dynamo with $k=|\alpha|/2\eta$ for which $B(t)=B_0 \times \exp \left\{\smallint_{t_0}^t[\alpha^2(t')/4\eta(t')]\mathrm{d}t' \right\}$.

Unfortunately, under the same conditions (for large $n_{\nu_e}=1.9\times 10^{37}\thinspace\text{cm}^{-3}$) the scale of the magnetic field occurs to be rather small, $\Lambda=k^{-1}\simeq \eta/|\alpha| =| \Pi_2|^{-1}\sim 1.25\times 10^{-3}\thinspace\text{cm}$. Here we use the fact that $| F|=2$, see figure~\ref{FTmu}(b). However, such a scale grows when the neutrino asymmetry diminishes due to a significant involvement of antineutrinos somewhere later at $t\leq  0.03-0.1\thinspace\text{s}$, $\Delta n_{\nu_e}=n_{\nu_e} - n_{\bar{\nu}_e}\to 0$ (see figure~11.3 in ref.~\cite{Raffelt}). It reaches the core radius $\Lambda\to R_0=10\thinspace\text{km}$, $k=| \Pi_2|=R_0^{-1}$, for the neutrino asymmetry density
\begin{equation}\label{largescale}
  n_{\nu_e} - n_{\bar{\nu}_e} =
  \frac{\pi}{R_0\alpha_\mathrm{em}G_\mathrm{F}\sqrt{2}| F|}
  \simeq
  5 \times 10^{27}\thinspace\text{cm}^{-3}.
\end{equation}
The magnetic diffusion time should be recalculated for this stage of  SN burst separately (the release of prompt $\nu_e$-burst due to the shock propagation and the following matter accretion, see in figure~11.1 in ref.~\cite{Raffelt}). However this task is beyond the scope of the present work.

The suggested mechanism of the $B$-field growth in a supernova driven by the electron neutrino asymmetry could lead to an additional amplification of a strong seed magnetic field ($B_0=10^{10}\div10^{12}\thinspace\text{G}$)  during the first second of a SN explosion when the asymmetry $n_{\nu_e} - n_{\bar{\nu}_e}\neq 0$ remains appreciable. Here a strong seed magnetic field can arise from a small magnetic field of a protostar, e.g., $B_\mathrm{proto} \sim 1 \div 10^2\thinspace\text{G}$, due to the conservation of the magnetic field flux, $B_0=B_\mathrm{proto}(R_\mathrm{proto}/R_0)^2$ , during the protostar collapse.  The question whether this new mechanism can explain the strongest magnetic field of observed magnetars ($B=10^{14}\div10^{15}$~G) deserves a separate study (see also in section \ref{COMPARISON}).

\subsection{Growth of primordial magnetic fields provided by the lower bound on neutrino asymmetries\label{EARLYUNI}}

In a hot plasma of the early universe the magnetic helicity parameter $\alpha$ in Faraday eq.~(\ref{Faraday}) reads as
\begin{equation}\label{alpha}
\alpha(T)=\frac{\Pi_2(T)}{\sigma_\mathrm{cond}(T)}=\frac{\alpha_\mathrm{em}G_\mathrm{F}\sqrt{2}T^2F(k_0/T)}{6\pi\sigma_c}\left[\xi_{\nu_e} - \xi_{\nu_{\mu}} - \xi_{\nu_{\tau}}\right],
\end{equation}
where we substituted the dimensionless neutrino asymmetries $\xi_{\nu_{\alpha}}=\mu_{\nu_{\alpha}}/T$ for the asymmetry densities $\Delta n_{\nu_{\alpha}}=\xi_{\nu_{\alpha}}T^3/6$ and used the hot plasma conductivity $\sigma_\mathrm{cond}=\sigma_cT$, with $\sigma_c\simeq 100$. The magnetic field evolution with the parameter $\alpha$ in eq.~(\ref{alpha}) obeys the {\it causal} scenario, where the magnetic field scale is less than the horizon, $\Lambda_\mathrm{B}\simeq \eta/| \alpha|<l_\mathrm{H}=H^{-1}$, if the sum of neutrino asymmetries $-2\sum_{\alpha}c^{(\mathrm{A})}_{\alpha}\xi_{\alpha}=\xi_{\nu_e} -\xi_{\nu_{\mu}} -\xi_{\nu_{\tau}}$ satisfies the inequality
\begin{equation}\label{inequality}
|\xi_{\nu_e} -\xi_{\nu_{\mu}} -\xi_{\nu_{\tau}}| >\frac{1.1\times 10^{-6}\sqrt{g^*/106.75}}{(T/\text{MeV})}.
\end{equation}
Here we take into account that $c^{(\mathrm{A})} = \mp 0.5$ (upper sign stays for electron neutrinos) is the SM axial coupling constant for $\nu e$ interaction corresponding to the difference $f^0_\mathrm{L} - f^0_\mathrm{R}$ in eq.~\eqref{deltaf}. In eq.~(\ref{alpha}) we use that $| F|\simeq 0.2$, which results from figure~\ref{FTmu}(a). Moreover we account for that $l_\mathrm{H}=M_0/T^2$, with $M_0=M_\mathrm{Pl}/1.66\sqrt{g^*}$, where $M_\mathrm{Pl}=1.2\times 10^{19}\thinspace\mathrm{GeV}$ is the Plank mass, $g^*=106.75$ is the number of relativistic degrees of freedom above the QCD phase transition, $T>T_\mathrm{QCD}\simeq 150\thinspace\text{MeV}$. Let us
remind that to get eq.~(\ref{inequality}) we applied the photon polarization term in eq.~(\ref{Fdef}) for ultrarelativistic leptons with $T\gg \max(m_e, \mu)$.

One can see that the inequality in eq.~(\ref{inequality}) does not contradict to the well-known BBN bounds
on the neutrino asymmetries at the lepton stage of the universe expansion corresponding to $g^*=10.75$, $|\xi_{\nu_{\alpha}}| <0.07$, (see ref.~\cite{Dolgov:2002ab}) and gives an additive (lower) bound on the neutrino asymmetry which supports the growth of CMF in our causal scenario. Here different flavors equilibrate due to neutrino oscillations before BBN, $\xi_{\nu_e}\sim \xi_{\nu_{\mu}}\sim \xi_{\nu_{\tau}}$, somewhere at the neutrino decoupling time $T=2-3\thinspace\text{MeV}$ \footnote{Neutrino oscillations are efficient at $T=3\thinspace\text{MeV}$, $E\simeq 3T$ even for the lowest (solar neutrino) mass difference $\Delta m^2_{\odot}=8\times 10^{-5}$ for which one gets the largest oscillation period $t_\mathrm{osc}=4\pi E/\Delta m^2\simeq 5\times 10^{-4}\thinspace\mathrm{s}$, which is much shorter than the Hubble time $H^{-1}(T=3\thinspace\text{MeV})\sim 0.1\thinspace\mathrm{s}$.}, accounting for all active neutrino flavors with the non-zero mixing angles (including $\sin^2 \theta_{13}=0.04$), see in ref.~\cite{Mangano:2011ip}.

We also obtain that the magnetic field diffusion time $t_\mathrm{diff}$ is bigger than the expansion time $\sim H^{-1}$, $t_\mathrm{diff}=\sigma_\mathrm{cond}/\Pi_2^2> M_0/T^2$, or ohmic losses are not danger, if the opposite inequality for neutrino asymmetries is valid,
\begin{equation}\label{inequality2}
|\xi_{\nu_e} -\xi_{\nu_{\mu}} -\xi_{\nu_{\tau}}| < \frac{12.7\times(g^*/106.75)^{1/4}}{[T/\mathrm{GeV}]^{3/2}}.
\end{equation}
Here just after the electroweak phase transition $T\leq T_\mathrm{EW}=100\thinspace\mathrm{GeV}$ the combined asymmetry in eq.~(\ref{inequality2})
seems to be resonable, $|\xi_{\nu_e} -\xi_{\nu_{\mu}} -\xi_{\nu_{\tau}}| < 0.013$, while at lower temperatures
$m_e\ll T\leq \mathcal{O}(\mathrm{GeV})$ the condition in eq.~(\ref{inequality2}) is obviously fulfilled and consistent with the BBN bound obtained in ref.~\cite{Dolgov:2002ab}.

\subsection{Comparison with the chiral magnetic mechanism in refs.~\cite{Boyarsky:2011uy,Yamamoto1,Yamamoto2} \label{COMPARISON}}

In ref.~\cite{Boyarsky:2011uy}, the magnetic helicity coefficient analogous to that in eq.~(\ref{alpha}) in our work,
\begin{equation}\label{BRFalpha}
\alpha (T)=\frac{\alpha_\mathrm{em}\Delta \mu (T)}{\pi\sigma_\mathrm{cond}(T)},
\end{equation}
is proportional to the magnetic chiral parameter $\Delta \mu=\mu_{e_\mathrm{L}} - \mu_{e_\mathrm{R}}$
where $\mu_{e_\mathrm{L}}$ ($\mu_{e_\mathrm{R}}$) are the left (right) electron chemical potentials. In QED plasma such a parameter arises due to the Adler anomaly  in external electromagnetic fields,
$\partial(j_\mathrm{L}^{\mu}- j_\mathrm{R}^{\mu})/\partial x^{\mu}=(2\alpha/\pi){\bf E}\cdot{\bf B}$, evolving in a self-consistent way with the magnetic field ${\bf B}$. However,
it tends to a small value $\Delta \mu/T\sim 10^{-6} - 10^{-7}$ for a small
wave number $10^{-10}\leq k/T\leq 3\times 10^{-9}$ at temperatures $T\geq 10\thinspace\text{MeV}$ (see figure~F.1 in ref.~\cite{Boyarsky:2011uy}) and vanishes later at all due to the chirality flip with the increasing rate $\Gamma_f\sim (m_e^2/T^2)$ in cooling universe, $n_{e_\mathrm{L}} - n_{e_\mathrm{R}}\to 0$. This is not the case for the helicity parameter given in eq.~(\ref{alpha}) based on neutrino asymmetries $\xi_{\nu_{\alpha}}$ for which there are no triangle anomalies in Maxwellian fields contrary to charged leptons
\footnote{Of course, triangle (Abelian) anomalies are possible for neutrinos in hypercharge fields $Y_{\mu}$ before electroweak phase transition (EWPT) since neutrinos interact with such fields \cite{DS,DS1}. In the present
work we study Maxwellian magnetic fields after EWPT.}. Moreover, after the neutrino decoupling and relic neutrino oscillations before BBN, there are no ways to change the equivalent asymmetries $\xi_{\nu_{\alpha}}=\text{const}\neq 0$, $\alpha=e,\mu,\tau$.

In ref.~\cite{Yamamoto2} one suggests a new mechanism for the production of strong magnetic fields in magnetars \cite{Duncan} based on the chiral instability for electrons with the difference of chemical potentials for right- and left-handed electrons, $\mu_5=\mu_\mathrm{R} - \mu_\mathrm{L}\neq 0$.  The chirality imbalance of electrons is produced via the same electron capture inside a core we considered above (urca-process), $p + e^{-}_\mathrm{L}\to n + \nu_\mathrm{L}^e$, where  the subscript L stands for left-handedness.

In ref.~\cite{Yamamoto2} the typical scales of wave number $k$ and vector potential $A$ relevant to such instability were obtained (see eq.~(27) there):
\begin{equation}
k \sim \alpha_\mathrm{em}\mu_5,\quad |\mathbf{A}|\sim \frac{\mu_5}{\alpha_\mathrm{em}},\quad\text{thus}\quad B\sim k|\mathbf{A}|\sim \mu_5^2,
\end{equation}
where an estimate $\mu_5=200\thinspace\text{MeV}$ gives huge $B_\mathrm{max}\sim \mu_5^2\sim 10^{18}\thinspace\text{G}$. The authors also show that instability proceeds faster than the danger chirality flip, $\Gamma_\mathrm{inst}=\alpha_\mathrm{em}^2\mu_5\gg \Gamma_\mathrm{flip}\sim \alpha_\mathrm{em}^2(m_e/\mu_5)^2\mu_5$, or the process $\mu_5\to 0$ due to collisions is negligible because $\mu_5\gg m_e$. In the proposed mechanism of the magnetic field amplification it remains unclear how a large-scale magnetic field is produced in this scenario since the magnetic field generated seems to be microscopic. Indeed, the scale $k^{-1}\sim \mathrm{MeV}^{-1}$ found in ref.~\cite{Yamamoto2} is much smaller than we obtain in section~\ref{SUPERNOV}. More realistic calculations using MHD for chiral plasma were suggested in ref.~\cite{Yamamoto2} to reach in future a definite conclusion.

\section{Conclusion\label{CONCL}}

In this work, we have shown a way to bridge between the photon self-energy (PSE) calculated by FTFT methods and its macroscopic consequences in magnetohydrodynamics (MHD). The latter is modified due to the appearance in SM of the parity violation  CS term $\Pi_2$ as a part of PSE. This leads to the new $\alpha$-helicity parameter in eq.~(\ref{helicity}) in Faraday equation that is {\it scalar} in modified dynamo theory instead of the well-known {\it pseudoscalar} $\alpha_\mathrm{MHD}\sim \langle \mathbf {v}\cdot(\nabla\times \mathbf{v})\rangle$ \cite{1983flma....3.....Z} which corresponds to the parity conservation in standard MHD.

While our astrophysical applications in the modified MHD are very preliminary, and at the present stage, for instance, does not involve Navier-Stokes equation for the fluid velocity $\mathbf{v}$, our calculations of PSE in electroweak plasma are exact and completed, at least, with the inclusion of $\nu l$ interaction in Fermi approximation.
Therefore we discuss below main steps in our calculations of PSE in quantum FTFT and compare the master CS term $\Pi_2$ we calculated here with results obtained by other authors.

We have started with the derivation of the exact propagator of a charged
lepton in the presence of the neutrino background. This propagator was used for the calculation of the loop
contribution to PSE. We have obtained the contributions of both virtual $l$'s, cf. eq.~\eqref{eq:Pivacexpl}, and plasma of particles having nonzero temperature and density, cf. eqs.~\eqref{Pi2Tgen} and~\eqref{J012}.

To obtain the plasma contribution to PSE the imaginary time perturbation theory has been used.
The expression for $\Pi_{2}^{(\nu l)}$ has been derived under the assumption that $k^2 < 4 m^2$, which means that no creation of $l\bar{l}$-pairs occurs in plasma. Indeed, as shown in ref.~\cite{Braaten:1993jw}, the plasmon decay is forbidden.

In eq.~(\ref{eq:lowtempclass}) we have derived the expression for
$\Pi_{2}^{(\nu l)}$ in case of a low temperature classical plasma of electrons.
We have obtained that $\Pi_{2}^{(\nu l)}\sim G_{\mathrm{F}}\alpha_{\mathrm{em}}$,
whereas $\Pi_2^{(\nu)}\sim G_{\mathrm{F}}\alpha_{\mathrm{em}}^{2}$.
Accounting for the additional constant factor, we get that $\Pi_{2}^{(\nu l)}$
is three orders of magnitude greater than $\Pi_2^{(\nu)}$.
Therefore one cannot neglect $\Pi_{2}^{(\nu l)}$ compared to
$\Pi_2^{(\nu)}$. Note that the plasma contribution to
the polarization tensor was overlooked in ref.~\cite{Mohanty:1997mr},
where the optical activity of the relic neutrino gas was studied and
only $\Pi_2^{(\nu)}$ was accounted for.

The computation of $\Pi_2$ for $e\bar{e}$ plasma and $\nu\bar{\nu}$ gas was made in ref.~\cite{NivSah05}, where, like in our work, the approximation of the Fermi interaction was used. In frames of the real time perturbation theory, the expression for the antisymmetric contribution to PSE was derived in ref.~\cite{NivSah05} on the basis of the modified electron's propagator. The cases of hot and degenerate relativistic plasmas were studied in ref.~\cite{NivSah05}. It was found that the contribution of virtual $l$'s to $\Pi_2$ is much smaller than $\Pi_2^{(\nu l)}$. However, using eq.~\eqref{eq:Pivacexpl} and the photon's dispersion laws found in sections~\ref{HOTRELPLAS} and~\ref{DEGRELPLAS}, we get that $\Pi_2^{(\nu)}$ can be comparable with $\Pi_2^{(\nu l)}$ in these media. The absolute values for the function $F$ in the static limit, obtained in ref.~\cite{NivSah05}, are different from these calculated in our work (see figures~\ref{FTmu}(a) and~\ref{FTmu}(b)). This discrepancy is because the radiative corrections to the electron mass in a hot and dense plasma, cf. eq.~\eqref{meff}, were not taken into account in ref.~\cite{NivSah05}.

Recently the two loop contribution to $\Pi_{2}$ was computed in ref.~\cite{BoyRucSha12}.
It was found there that $\Pi_{2}=\tfrac{\alpha_{\mathrm{em}}}{2\pi} \tfrac{4G_{\mathrm{F}}}{\sqrt{2}}\left(\sum_{i}c_{Li}L_{i}+c_\mathrm{B}B\right)$,
where \textbf{$L_{i}$} and $B$ are conserved charges of all kinds
of leptons, including neutrinos, and baryons, $c_{Li}$ and $c_\mathrm{B}$
are the constant factors. That value of $\Pi_{2}$ is nonvanishing for the background medium with a nonzero asymmetry of the $\nu\bar{\nu}$ gas, $L_{\nu\text{'s}}\neq0$, even if all charged leptons and quarks are virtual
particles, $L_{i\neq\nu\text{'s}}=0$ and $B=0$.
It should be noted that the value of $\Pi_{2}$ in ref.~\cite{BoyRucSha12} depends on the type of plasma where a photon propagates with conservation of global charges $B/3 - L_a=\text{const}$ appropriate for a given medium, $a=e,\mu,\tau$.

We have explicitly demonstrated that the static value of $\Pi_{2}$ can have different values depending on what kind of plasma of charged leptons is studied. In the lowest order over $G_\mathrm{F}$ and in the case of a classical electron plasma with low temperature and density {\it the total} $\Pi_{2}=\Pi_2^{(\nu)} + \Pi_2^{(\nu l)} \to 0$ at $k^2 \to 0$ because in vacuum filled by neutrinos (antineutrinos) there are no real electrons, $n_e=0$. This result is in agreement with ref.~\cite{Gel61}. However, in the same approximation, $\sim G_\mathrm{F}$, in the case of hot relativistic and degenerate relativistic plasmas coexisting with neutrino background $\Pi_{2} \neq 0$ at $k_0 \to 0$. Nevertheless, the values of $\Pi_2(0)$ are different for these plasmas (see eq.~\eqref{Fdef} and figure~\ref{FTmu}).

Therefore our study of the parity violating effects in PSE generalizes the results of ref.~\cite{BoyRucSha12} since our expression for $\Pi_2$ exactly accounts for charged lepton plasma characteristics, like $T$ and $\mu$, the lepton mass, which should not be omitted in hot and dense matter (see, e.g., ref.~\cite{Kli82}), as well as the photon dispersion law in this matter.
Moreover, unlike ref.~\cite{BoyRucSha12}, our method of calculations allows  to reproduce the value of $\Pi_2=0$ corresponding to the case of $\nu\bar{\nu}$ gas with the nonzero asymmetry, $n_\nu - n_{\bar{\nu}} \neq 0$, and purely virtual charged leptons, $T=0$ and $\mu = 0$.

In section~\ref{APPL} we have applied our main result in eq.~(\ref{Fdef}) to the magnetic field evolution in both (i) ultrarelativistic degenerate electron gas of a supernova during its neutrino burst and (ii) hot plasma of early universe at $T\gg m_e$. In both cases one finds a possibility of the magnetic field growth driven by the neutrino asymmetries and avoiding the magnetic diffusion. Of course, a seed field $B_0$ should be assumed at the initial time instance $t_0$ which then evolves through the $\alpha$-dynamo mechanism as $B(k,t)=B_0\exp [\int_{t_0}^t(|\alpha| k - \eta k^2)dt^{'}]$ for a given wave number $k$ and the magnetic field scale $\Lambda_\mathrm{B} = k^{-1}$. The question whether the new $B$-field amplification mechanism in a supernova driven by a nonzero $(n_{\nu_e} - n_{\bar{\nu}_e})$-asymmetry could lead to an explanation of strongest magnetic fields observed in magnetars deserves a separate study.

We have found an interesting lower bound in eq.~(\ref{inequality}) on the combined neutrino asymmetry providing the CMF growth in a hot plasma of the early universe which is consistent with the well-known BBN (upper) bound on the electron neutrino asymmetry \cite{Dolgov:2002ab}.
We suppose that the new mechanism suggested in section~\ref{EARLYUNI} for a $B$-field growth driven by neutrino asymmetries in a hot plasma  can be more productive in comparison with that involving the chiral electron asymmetry $\sim (\mu_\mathrm{R}-\mu_\mathrm{L})$~\cite{Boyarsky:2011uy}. For the latter the chirality flip due to collisions with the rate $\Gamma_f\sim m_e^2/T^2$, washes out the corresponding $\alpha$-magnetic helicity parameter stronger and stronger in cooling universe. In contrast to that the neutrino asymmetries $\sim\xi_{\alpha}$ equilibrate before BBN due to neutrino oscillations being conserved even after neutrino decoupling if the total lepton number is conserved.

\acknowledgments

We are thankful to O.~Ruchayskiy for helpful comments, to M.~Shaposhnikov for communications and D.~Sokoloff for some discussions. M.D. acknowledges FAPESP (Brazil) for a grant, as well as Y.~Kivshar for the hospitality at the ANU where a part of the work was made.

\appendix

\section{Propagator of a charged lepton interacting with a neutrino gas\label{PROP}}

In this Appendix we briefly describe the interaction between charged leptons and neutrinos in frames of the Fermi theory. Then we derive the exact propagator of a charged lepton in the presence of the neutrino-antineutrino gas.

The evolution of a charged lepton $l$, represented as a bispinor
$\psi$, interacting with the $\nu\bar{\nu}$ gas, is described by the following Dirac equation \cite{PivStu05}:
\begin{equation}\label{eq:Direqpsi}
  \left[
    \mathrm{i}\gamma^{\mu}\partial_{\mu}-
    \gamma_{\mu}
    \left(
      f^{\mu}_{\mathrm{L}} P_{\mathrm{L}} +
      f^{\mu}_{\mathrm{R}} P_{\mathrm{R}}
    \right) - m
  \right]\psi=0,
\end{equation}
where $m$ is the mass of $l$, $\gamma^{\mu}=(\gamma^{0},\bm{\gamma})$
are the Dirac matrices, $P_{\mathrm{L,R}}=(1\mp\gamma^{5})/2$ are the
chiral projectors, $\gamma^{5}=\mathrm{i}\gamma^{0}\gamma^{1}\gamma^{2}\gamma^{3}$.

The $l\nu$ interaction in eq.~\eqref{eq:Direqpsi} can be described in the mean field approximation via the external neutrino macroscopic currents $f^\mu_{\mathrm{L,R}} = (f^0_{\mathrm{L,R}},\mathbf{f}_{\mathrm{L,R}})$. To find the explicit form of $f^\mu_{\mathrm{L,R}}$ we shall consider
a background matter consisting of the $\nu_\alpha\bar{\nu}_\alpha$ gas, $\alpha=e,\mu,\tau$, and identify $l$ with an electron. The effective Lagrangian for the $\nu e$-interaction
has the form~\cite{GiuKim07},
\begin{equation}\label{Leff}
  \mathcal{L}_{\mathrm{eff}} = -\sqrt{2}G_{\mathrm{F}}
  \sum_{\alpha}
  \bar{\nu}_\alpha \gamma^{\mu} (1-\gamma^5) \nu_\alpha \cdot
  \bar{\psi}\gamma_{\mu}
  \left(
    a_{\mathrm{L}}^{(\alpha)} P_{\mathrm{L}} + a_{\mathrm{R}}^{(\alpha)} P_{\mathrm{R}}
  \right)
  \psi,
\end{equation}
where $G_{\mathrm{F}} \approx 1.17 \times 10^{-5}\thinspace\text{GeV}^{-2}$ is the Fermi constant,
\begin{equation}\label{aLlamaRlam}
  a_{\mathrm{L}}^{(\alpha)} = \delta_{\alpha,e} + \sin^2\theta_\mathrm{W} -1/2,
  \quad
  a_{\mathrm{R}}^{(\alpha)} = \sin^2\theta_\mathrm{W},
\end{equation}
and $\theta_\mathrm{W}$ is the Weinberg angle. The symbol $\delta_{e,\alpha}$ in eq.~\eqref{aLlamaRlam} equals to one if $\alpha = e$ and to zero otherwise. To derive eq.~\eqref{Leff} we use the Fierz transformation and take into account that $\psi$ and $\nu_\alpha$ are anticommuting operator valued spinors.

For applications above we calculate $\Pi_2$ in the case of the isotropic $\nu\bar{\nu}$ gas. It means that, in averaging over the neutrino ensemble, the only nonzero quantity is $\langle \bar{\nu}_\alpha \gamma^{0} (1-\gamma^5) \nu_\alpha \rangle$. Nowadays it is generally believed that neutrinos possess nonzero masses leading to observed neutrino oscillations in numerous underground experiments. In section~\ref{APPL}, where we discuss applications of our results, we consider for simplicity the case of the $\nu\bar{\nu}$ gas consisting of massless neutrinos hence neglecting neutrino oscillations influence the magnetic field generation. In this situation
$\langle \bar{\nu}_\alpha \gamma^0 (1-\gamma^5) \nu_\alpha \rangle = 2 \Delta n_{\nu_\alpha}$, where $\Delta n_{\nu_\alpha} = n_{\nu_\alpha} - n_{\bar{\nu}_\alpha}$ and
\begin{equation}\label{nnugen}
  n_{\nu_\alpha,\bar{\nu}_\alpha} =
  \int\frac{\mathrm{d}^{3}p}{(2\pi)^{3}}
  \left[
    \exp
    \left(
      \frac{|\mathbf{p}| \mp \mu_{\nu_\alpha}}{T_{\nu_\alpha}}
    \right)+1
  \right]^{-1},
\end{equation}
are the number densities of neutrinos and antineutrinos. In eq.~\eqref{nnugen} $T_{\nu_\alpha}$ and $\mu_{\nu_\alpha}$ are the temperature and the chemical
potential of the $\alpha$ component of the $\nu\bar{\nu}$ gas.
Using eqs.~\eqref{eq:Direqpsi}-\eqref{nnugen} we get that $\mathbf{f_{\mathrm{L,R}}} = 0$ and
\begin{equation}\label{f0LR}
  f^0_\mathrm{L} = 2\sqrt{2} G_{\mathrm{F}}
  \left[
    \Delta n_{\nu_e} +
    (\sin^2\theta_\mathrm{W} -1/2)
    \sum_\alpha \Delta n_{\nu_\alpha}
  \right],
  \quad
  f^0_\mathrm{R} = 2\sqrt{2} G_{\mathrm{F}}
  \sin^2\theta_\mathrm{W}
  \sum_\alpha \Delta n_{\nu_\alpha}.
\end{equation}
%


Basing on eq.~(\ref{eq:Direqpsi}) one finds that the Fourier transform
of the Green function $S(x)=\int\frac{\mathrm{d}^{4}p}{(2\pi)^{4}}e^{-\mathrm{i}px}S(p)$
of the field $\psi$ satisfies the equation,
\begin{equation}\label{eq:DireqS}
  \left[
    \gamma^{\mu}p_{\mu} - \gamma_{\mu}
    \left(
      f^{\mu}_{\mathrm{L}} P_{\mathrm{L}} +
      f^{\mu}_{\mathrm{R}} P_{\mathrm{R}}
    \right) - m
  \right]S(p)=1.
\end{equation}
Using the results of ref.~\cite{PivStu05} one can derive the expression
for $S$,
\begin{equation}\label{eq:Sgen}
  S(p) =
  \frac{
  \left[
    P^{2}-m^{2}-
    (f_\mathrm{L} - f_\mathrm{R})^2/4 +
    \mathrm{i}\sigma_{\alpha\beta}\gamma^{5}P^{\alpha}
    (f^{\beta}_\mathrm{L} - f^{\beta}_\mathrm{R})
  \right]
  \left[
    \gamma^{\mu}P_{\mu}+m+\gamma_{\mu}\gamma^{5}
    (f^{\mu}_\mathrm{L} - f^{\mu}_\mathrm{R})/2
  \right]}
  {
  \left[
    P^{2}-m^{2}-(f_\mathrm{L} - f_\mathrm{R})^2/4
  \right]^{2} +
  \left[
    P^{2}(f_\mathrm{L} - f_\mathrm{R})^2 -
    (f^{\mu}_\mathrm{L} - f^{\mu}_\mathrm{R} \cdot P_\mu)^{2}
  \right]},
\end{equation}
where $P^{\mu}=p^{\mu}-(f^{\mu}_\mathrm{L} + f^{\mu}_\mathrm{R})/2$ is the canonical momentum and $\sigma_{\mu\nu}=\tfrac{\mathrm{i}}{2}\left[\gamma_{\mu},\gamma_{\nu}\right]_{-}$.

Note that the denominator $\mathcal{D}$ of $S$ in eq.~\eqref{eq:Sgen} is the fourth order polynomial with respect to $p^0$. In general the equation $\mathcal{D}=0$ has four different roots $p^0_{1,\dots,4}$, which determine the poles of $S$. The Green function in eq.~\eqref{eq:Sgen} corresponds to a propagator if we bypass its poles in the complex plane in a standard manner: positive roots should be added a small negative imaginary contribution $-\mathrm{i}0$ and negative ones acquire $+\mathrm{i}0$.

It is convenient to represent $S$ as a series $S=S_{0}+S_{1}+\dotsb$,
keeping only the terms linear in $f^{\mu}_\mathrm{L,R}$ which contain $\gamma^{5}$
since they are responsible for the parity violation. The explicit
forms of $S_{0}$ and $S_{1}$ are
\begin{align}\label{eq:Sexpand}
  S_{0} = & \frac{\gamma^{\mu}P_{\mu}+m}{P^{2}-m^{2}},
  \notag
  \\
  S_{1} = & \frac{1}{P^{2}-m^{2}}
  \left[
    \frac{\mathrm{i}\sigma_{\alpha\beta}\gamma^{5}P^{\alpha}
    (f^{\beta}_\mathrm{L} - f^{\beta}_\mathrm{R})
    (\gamma^{\mu}P_{\mu}+m)}
    {P^{2}-m^{2}} +
    \frac{1}{2}\gamma_{\mu}\gamma^{5}
    (f^{\mu}_\mathrm{L} - f^{\mu}_\mathrm{R})
  \right],
\end{align}
where we should take into account that $m^2 \to m^2 - \mathrm{i}0$ in the denominators.

\section{Dimensional regularization\label{DIMREG}}

The dimensional regularization is introduced in the following way~\cite{Aok82}:
\begin{equation}
  \int\frac{\mathrm{d}^{4}p}{(2\pi)^{4}} \to
  \int\frac{\mathrm{d}^{N}p}{(2\pi)^{N}},
\end{equation}
where $N = 4 - 2\varepsilon$ and $\varepsilon \to 0$. The convolution of the metric tensor reads $g_{\mu\nu}g^{\mu\nu} = N$.

Some useful momentum integrals have the form~\cite{BogShi80},
\begin{align}\label{intdimreg}
  \mathrm{i} \int \frac{\mathrm{d}^{N}p}{(2\pi)^{N}}
  \frac{1}{\left[p^{2}-M^{2}\right]^{2}} & =
  - \frac{1}{16\pi^{2}}
  \left[
    \frac{4\pi\lambda^{2}}{M^{2}}
  \right]^{\varepsilon}\Gamma(\varepsilon),
  \notag
  \\
  \mathrm{i} \int\frac{\mathrm{d}^{N}p}{(2\pi)^{N}}
  \frac{1}{\left[p^{2}-M^{2}\right]^{3}} & =
  \frac{1}{32\pi^{2}}
  \left[
    \frac{4\pi\lambda^{2}}{M^{2}}
  \right]^{\varepsilon}
  \frac{\Gamma(1+\varepsilon)}{M^{2}},
  \notag
  \\
  \mathrm{i} \int\frac{\mathrm{d}^{N}p}{(2\pi)^{N}}
  \frac{p_{\mu}p_{\nu}}{\left[p^{2}-M^{2}\right]^{2}} & =
  -\frac{1}{16\pi^{2}}
  \left[
    \frac{4\pi\lambda^{2}}{M^{2}}
  \right]^{\varepsilon}
  g_{\mu\nu} \frac{M^{2}\Gamma(\varepsilon)}{2(1-\varepsilon)},
  \notag
  \\
  \mathrm{i} \int\frac{\mathrm{d}^{N}p}{(2\pi)^{N}}
  \frac{p_{\mu}p_{\nu}}{\left[p^{2}-M^{2}\right]^{3}} & =
  -\frac{1}{64\pi^{2}}
  \left[
    \frac{4\pi\lambda^{2}}{M^{2}}
  \right]^{\varepsilon}
  g_{\mu\nu}\Gamma(\varepsilon),
\end{align}
where $\Gamma(z)$ is the Euler Gamma function and $\lambda$ is the parameter having the dimension of mass.

\section{Technique for the integrals calculation\label{FTFTINT}}

In this appendix we give an example of the calculation of one of the integrals which contribute to eqs.~\eqref{Pi2Tgen} and~\eqref{J012}. To account for the plasma effects we use the technique for the summation over Matsubara frequencies.

Let us consider the Feynman integral in Minkowski space
\begin{equation}\label{Iex}
  I = \mathrm{i} \int \frac{\mathrm{d}^4 p}{(2\pi)^4} \frac{1}{[p^2-M^2]^2}
\end{equation}
where $M^2$ is a positive real parameter. To get the analog of $I$ in the presence of a fermionic plasma, we should transform eq.~\eqref{Iex} to (see eq.~\eqref{Mfintr} and ref.~\cite{KapGal06})
\begin{equation}\label{IexMf}
  I = T \sum_n \int \frac{\mathrm{d}^3 p}{(2\pi)^3} \frac{1}{[p^2-M^2]^2},
\end{equation}
where $p^2 = [(2n+1)\pi T\mathrm{i}+\mu]^2 - \mathbf{p}^2$, $T$ is the plasma temperature, and $\mu$ is the plasma chemical potential.

\begin{figure}
  \centering
  \includegraphics[scale=.4]{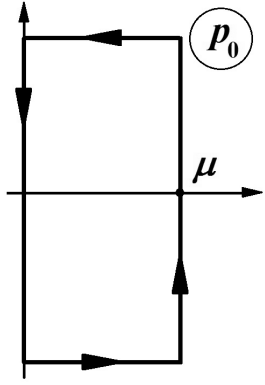}
  \caption{The contour $C$ for the integration in eq.~(\ref{IexsumMf}).\label{contC}}
\end{figure}

The summation over the Matsubara frequencies in eq.~\eqref{IexMf} for fermions can be replaced by the integration in a complex plane~\cite{KapGal06},
\begin{align}\label{IexsumMf}
  I = &
  \bigg[
    - \frac{1}{2\pi\mathrm{i}}
    \int_{-\mathrm{i}\infty+\mu+\epsilon}^{+\mathrm{i}\infty+\mu+\epsilon}
    \frac{\mathrm{d}p^{0}}{\exp[\beta(p^{0} - \mu)]+1}
    \notag
    \\
    & -
    \frac{1}{2\pi\mathrm{i}}
    \int_{-\mathrm{i}\infty+\mu-\epsilon}^{+\mathrm{i}\infty+\mu-\epsilon}
    \frac{\mathrm{d}p^{0}}{\exp[\beta(\mu - p^{0})]+1}
    \notag
    \\
    & +
    \frac{1}{2 \pi \mathrm{i}}\oint_{C}dp^{0}
  \bigg]
  \times
  \int \frac{\mathrm{d}^3 p}{(2\pi)^3}
  \frac{1}{[p_0^2-\mathcal{E}_\mathbf{p}^2]^2},
\end{align}
where $\mathcal{E}_\mathbf{p} = \sqrt{\mathbf{p}^2 + M^2}$, $\beta$ is the reciprocal of the temperature, and the contour $C$ is shown in figure~\ref{contC}.
Note that in eq.~\eqref{IexsumMf} we keep only the terms which depend on $T$ and $\mu$.

Using the expression for the residue in the pole of order $n=2$,
\begin{equation}
  \mathrm{Res}(f,c) = \frac{1}{(n-1)!}
  \lim_{z \to c} \frac{\mathrm{d}^{n-1}}{\mathrm{d}z^{n-1}}
  \left[
    (z-c)^n f(z)
  \right],
\end{equation}
one can calculate the integrals in eq.~\eqref{IexsumMf},
\begin{align}\label{Iintegr}
  \int_{-\mathrm{i}\infty+\mu+\epsilon}^{+\mathrm{i}\infty+\mu+\epsilon}
  &
  \frac{\mathrm{d}p^{0}}{\exp[\beta(p^{0} - \mu)]+1}
  \frac{1}{[p_0^2-\mathcal{E}_\mathbf{p}^2]^2}
  \notag
  \\
  & =
  \frac{\pi \mathrm{i}
  \left[
    \theta(\mathcal{E}_\mathbf{p}-\mu)\theta(\mu)+\theta(-\mu)
  \right]
  }
  {2\mathcal{E}_\mathbf{p}^{2}
  \left(
    \exp[\beta(\mathcal{E}_\mathbf{p}-\mu)]+1
  \right)
  }
  \left\{
    \frac{\beta}{\exp[\beta(\mu-\mathcal{E}_\mathbf{p})]+1} +
    \frac{1}{\mathcal{E}_\mathbf{p}}
  \right\}
  \notag
  \\
  & +
  \frac{\pi \mathrm{i}\theta(-\mathcal{E}_\mathbf{p}-\mu)\theta(-\mu)}
  {2\mathcal{E}_\mathbf{p}^{2}
  \left(
    \exp[-\beta(\mathcal{E}_\mathbf{p}+\mu)]+1
  \right)
  }
  \left\{
    \frac{\beta}{\exp[\beta(\mathcal{E}_\mathbf{p}+\mu)]+1} -
    \frac{1}{\mathcal{E}_\mathbf{p}}
  \right\},
  \notag
  \\
  \int_{-\mathrm{i}\infty+\mu-\epsilon}^{+\mathrm{i}\infty+\mu-\epsilon}
  &
  \frac{\mathrm{d}p^{0}}{\exp[\beta(\mu - p^{0})]+1}
  \frac{1}{[p_0^2-\mathcal{E}_\mathbf{p}^2]^2}
  \notag
  \\
  & =
  \frac{\pi \mathrm{i}
  \left[
    \theta(\mathcal{E}_\mathbf{p}+\mu)\theta(-\mu)+\theta(\mu)
  \right]
  }
  {2\mathcal{E}_\mathbf{p}^{2}
  \left(
    \exp[\beta(\mathcal{E}_\mathbf{p}+\mu)]+1
  \right)
  }
  \left\{
    \frac{\beta}{\exp[-\beta(\mathcal{E}_\mathbf{p}+\mu)]+1} +
    \frac{1}{\mathcal{E}_\mathbf{p}}
  \right\}
  \notag
  \\
  & +
  \frac{\pi \mathrm{i}\theta(\mu-\mathcal{E}_\mathbf{p})\theta(\mu)}
  {2\mathcal{E}_\mathbf{p}^{2}
  \left(
    \exp[\beta(\mu-\mathcal{E}_\mathbf{p})]+1
  \right)
  }
  \left\{
    \frac{\beta}{\exp[\beta(\mathcal{E}_\mathbf{p}-\mu)]+1} -
    \frac{1}{\mathcal{E}_\mathbf{p}}
  \right\},
  \notag
  \\
  \oint_{C} &
  \frac{\mathrm{d}p^{0}}{[p_0^2-\mathcal{E}_\mathbf{p}^2]^2} =
  -\frac{\pi \mathrm{i}}{2\mathcal{E}_\mathbf{p}^{3}}
  \left[
    \theta(\mu)\theta(\mu-\mathcal{E}_\mathbf{p}) +
    \theta(-\mu)\theta(-\mathcal{E}_\mathbf{p}-\mu)
  \right],
\end{align}
where $\theta(z)$ is the Heaviside step function.

Finally, using eqs.~\eqref{IexsumMf} and~\eqref{Iintegr} we obtain for $I$,
\begin{align}
  I = & - \frac{1}{4}
  \int\frac{\mathrm{d}^{3}p}{(2\pi)^{3}}
  \frac{1}{\mathcal{E}_\mathbf{p}^{3}}
  \bigg\{
    \frac{1}{\exp[\beta(\mathcal{E}_\mathbf{p}-\mu)]+1} +
    \frac{1}{\exp[\beta(\mathcal{E}_\mathbf{p}+\mu)]+1}
    \notag
    \\
    & +
    \frac{\beta\mathcal{E}_\mathbf{p}}{2}
    \left[
      \frac{1}{\cosh[\beta(\mathcal{E}_\mathbf{p}-\mu)]+1} +
      \frac{1}{\cosh[\beta(\mathcal{E}_\mathbf{p}+\mu)]+1}
    \right]
  \bigg\}.
\end{align}
All the temperature and chemical potential dependent integrals which lead to eqs.~\eqref{Pi2Tgen} and~\eqref{J012} can be can be calculated in a similar manner.

\section{General system of evolution equations for the spectra of the helicity density and the magnetic energy density \label{SYSTEM}}

We derive this system for the case of helical magnetic fields in the early universe starting from the Faraday eq.~(\ref{Faraday}) to show how its solution in eq.~(\ref{Balpdyn})  can be obtained for maximum helical magnetic field . The analogous system can be presented for the case of magnetic fields in a supernova. The method is similar to that in Appendix E in ref.~\cite{Boyarsky:2011uy} for CMF and in ref.~\cite{SSS} for the helical hypermagnetic field. The magnetic helicity $H=\int \mathrm{d}^3 x ({\bf A}\cdot{\bf B})$ is an inviscid invariant of motion in the contemporary universe. The corresponding conservation law $\mathrm{d}H/\mathrm{d}t=0$ in an ideal plasma ($\sigma_\mathrm{cond}\to \infty$) severely constraints the magnetic field generation by the dynamo process and its further evolution.

Multiplying eq.~(\ref{Faraday}) by the
corresponding vector potential and adding the analogous construction
produced by the evolution equation governing the vector potential
(multiplied by a magnetic field), after the integration over space we
get the evolution equation for the magnetic helicity $H =\int \mathrm{d}^3 x  ({\bf A}\cdot{\bf B})$,
\begin{align}\label{helicityeq}
  \frac{{\rm d}H}{{\rm d}t} = &
  -2\int_V({\bf E}\cdot{\bf B})\mathrm{d}^3 x  - \oint [A_0{\bf B}
  +{\bf E}\times {\bf A}]\mathrm{d}^2 S
  \nonumber
  \\
  & =
  -2\beta (t)\int \mathrm{d}^3 x
  (\nabla\times {\bf B})\cdot {\bf B} +
  2\alpha(t)\int \mathrm{d}^3 x B^2(t),
\end{align}
where $\alpha$ is given by eq.~(\ref{helicity}), $\beta=(\sigma_\mathrm{cond})^{-1}$ is the magnetic diffusion coefficient given by the plasma conductivity $\sigma_\mathrm{cond}$. We changed here the notation $\eta=(\sigma_\mathrm{cond})^{-1}$ used above in section~\ref{APPL} to avoid its confusion with the standard notation for the conformal time (see below).

The surface integral $\oint(\dots)$ was omitted in the last line in eq.~(\ref{helicityeq}) since
electromagnetic fields vanish at infinity.
Let us change physical variables to the conformal ones using the conformal time $\eta=M_0/T$, $M_0=M_\mathrm{Pl}/1.66\sqrt{g^*}$, where $M_\mathrm{Pl}=1.2\times 10^{19}\thinspace\text{GeV}$ is the Plank mass, $g^*=106.75$ is the effective number of relativistic degrees of freedom.

In Friedmann-Robertson-Walker metric $\mathrm{d}s^2=a^2(\eta)(\mathrm{d}\eta^2 - \mathrm{d}\tilde{{\bf x}}^2)$, using the definitions $a=T^{-1}$, where $a_0=1$ at the present temperature $T_\mathrm{now}$, and $\mathrm{d}\eta=\mathrm{d}t/a(t)$ , we input the following notations: $\tilde{k}=ka=\mathrm{const}$ is the conformal momentum (giving a red shift for the physical one, $k\sim T=T_\mathrm{now}[1 + z]$), $\tilde{\Pi}_2(\eta)=a\Pi_2 =\Pi_2/T$ is the dimensional CS term in PSE proportional to the neutrino asymmetry $\Delta n_{\nu}$ and changing over time, $\tilde{{\bf B}}=a^2{\bf B}$, and $\tilde{\bf A}=a{\bf A}$ are the conformal dimensionless counterparts of the magnetic field and the vector potential correspondingly.

It is suitable to rewrite eq.~(\ref{helicityeq}) using the conformal coordinate $\tilde{{\bf x}}={\bf x}/a$ for the Fourier components
of the helicity density,
$\tilde{h}(\eta)\equiv \int (\tilde{\bf A}\cdot\tilde{\bf B})\mathrm{d}^3 x /V=\int \mathrm{d}\tilde{k} \tilde{h}(\tilde{k},\eta)$, and the magnetic energy density
$\tilde{\rho}_\mathrm{B}(\eta)=\tilde{B}^2(\eta)/2=\int \mathrm{d}\tilde{k}\tilde{\rho}_\mathrm{B}(\tilde{k},\eta)$ defined as their spectra,
\begin{eqnarray}\label{Fourier}
&&\tilde{h}(\tilde{k},\eta)=\frac{\tilde{k}^2a^3}{2\pi^2 V}\tilde{{\bf A}}(\tilde{k},\eta)\cdot\tilde{{\bf B}}^*(\tilde{k},\eta),\nonumber\\&&\tilde{\rho}_\mathrm{B}(\tilde{k},\eta)=\frac{\tilde{k}^2a^3}{4\pi^2V}\tilde{{\bf B}}(\tilde{k},\eta)\cdot\tilde{{\bf B}}^*(\tilde{k},\eta).
\end{eqnarray}
The obtained expressions allow us to calculate integrals $\int \mathrm{d}^3 x (...)/V$ in eq.~(\ref{helicityeq}) as well as in Faraday eq.~(\ref{Faraday}) multiplied by ${\bf B}^*$ and added with its complex conjugated product ${\bf B}\partial_t{\bf B}^*=\alpha\mathbf{B}(\nabla\times \mathbf{B}^*) + \beta\mathbf{B}\nabla^2\mathbf{B}^*$  to get $\partial_t\rho_B=\partial_t(\mathbf{B^*\mathbf{B}})/2$ and then to derive both the evolution equation for the magnetic helicity density and the magnetic energy density spectra.

The general system of the evolution equations for the spectra of the helicity density $\tilde{h}(\tilde{k},\eta)$ and the energy density $\tilde{\rho}_\mathrm{B}(\tilde{k},\eta)$ obeying the inequality $\tilde{\rho}_\mathrm{B}(\tilde{k},\eta)\geq \tilde{k}\tilde{h}(\tilde{k},\eta)/2$ \cite{Biskamp} has the following form in conformal variables:
\begin{eqnarray}\label{general}
  &&\frac{\mathrm{d}\tilde{h}(\tilde{k},\eta)}{\mathrm{d}\eta} =
  -\frac{2\tilde{k}^2}{\sigma_c}\tilde{h}(\tilde{k},\eta) +
  \left(\frac{4\tilde{\Pi}_2}{\sigma_c}\right)\tilde{\rho}_\mathrm{B}(\tilde{k},\eta)
  \nonumber\\&&
  \frac{\mathrm{d}\tilde{\rho}_\mathrm{B}(\tilde{k},\eta)}{\mathrm{d}\eta}=
  -\frac{2\tilde{k}^2}{\sigma_c}\tilde{\rho}_\mathrm{B}(\tilde{k},\eta)+
  \left(\frac{\tilde{\Pi}_2}{\sigma_c}\right)\tilde{k}^2 \tilde{h}(\tilde{k},\eta),
\end{eqnarray}
where $\sigma_c=\sigma_\mathrm{cond}a=\sigma_\mathrm{cond}/T\approx 100$ is the dimensionless plasma conductivity and $\Pi_2$ is the CS term in PSE given by eq.~(\ref{Fdef}).

It would be interesting in future to study, using eq.~(\ref{general}), how the initial non-helical field, $\tilde{h}(\tilde{k},\eta_0)=0$, evolves in the presence of a non-zero initial energy spectrum for which $[\mathrm{d}\tilde{h}(\tilde{k},\eta)/\mathrm{d}\eta]_{\eta=\eta_0}=[4\tilde{\Pi}_2(\eta_0)/\sigma_c]\tilde{\rho}_\mathrm{B}(\tilde{k},\eta_0)\neq 0$.

For the particular case of the maximum helicity
\begin{equation}\label{maximum}\tilde{h}(\tilde{k},\eta)=2\tilde{\rho}_\mathrm{B}(\tilde{k},\eta)/\tilde{k},
\end{equation}
the system in eq.~(\ref{general}) reads as the single equation,
\begin{equation}\label{conformal}
  \frac{\mathrm{d}\tilde{h}(\tilde{k},\eta)}{\mathrm{d}\eta}=
  -\frac{2\tilde{k}^2\tilde{h}(\tilde{k},\eta)}{\sigma_c}+
  \left(\frac{2\tilde{\Pi}_2\tilde{k}}{\sigma_c}\right)
  \tilde{h}(\tilde{k},\eta).
\end{equation}
Such a choice of the fully helical magnetic field in eq.~(\ref{maximum}) allows one to get the simple differential eq.~(\ref{conformal})
and provides an efficient inverse cascade for turbulent Maxwellian magnetic fields.

The solution of eq.~(\ref{conformal}) takes the form (compare to eq.~(8) in ref.~\cite{Boyarsky:2011uy}):
\begin{equation}\label{helicitysolution}
\tilde{h}(\tilde{k},\eta)=\tilde{h}^{(0)}(\tilde{k},\eta_0)\exp \left(\frac{2\tilde{k}}{\sigma_c}\left[\int_{\eta_0}^{\eta}
\tilde{\Pi}_2(\eta')\mathrm{d}\eta' - \tilde{k}(\eta - \eta_0)\right]\right).
\end{equation}
The spectrum of the dimensionless helicity density $\tilde{h}(\tilde{k},\eta)=a^3h(\tilde{k},\eta)$ can be rewritten in compact form as
 \begin{equation}\label{conformsolution}
\tilde{h}(\tilde{k},\eta)\equiv\frac{h(\tilde{k},\eta)}{T^3}= \tilde{h}^{(0)}(\tilde{k},\eta_0)\exp \left[A(\eta)\tilde{k} -B(\eta)\tilde{k}^2\right],
 \end{equation}
 where the initial spectrum $\tilde{h}^{(0)}(\tilde{k},\eta_0)=h(\tilde{k},\eta_0)/T_0^3$ corresponds in our scenario to an initial moment, and we used notations taken from eq.~(\ref{helicitysolution})
 \begin{equation}\label{parameter} A(\eta)=\frac{2}{\sigma_c}\int_{\eta_0}^{\eta}\tilde{\Pi}_2(\eta')d\eta',~~~~B(\eta)=\frac{2}{\sigma_c}(\eta - \eta_0).
 \end{equation}

In the ideal plasma limit, $\sigma_c\to \infty$, we get from eq.~(\ref{conformsolution}) the standard conservation of the helicity density, $\mathrm{d}\tilde{h}/\mathrm{d}\eta =0$ or $\tilde{h}=\text{const}$, with the conformal scaling $h(\eta)=(\eta_0/\eta)^3h(\eta_0)$.

Using  the connection of spectra of the energy density $\tilde{\rho}_B(\tilde{k},\eta)$ and the maximum helicity $\tilde{h}(\tilde{k},\eta)$ in eq.~(\ref{maximum}) one can easily find the corresponding solution of Faraday eq.~(\ref{Faraday}) coming from eq.~(\ref{helicitysolution}),
\begin{eqnarray}\label{mono}
\tilde{B}(\tilde{k},\eta)=&&\tilde{B}_0(\tilde{k},\eta_0)\exp \left(\frac{\tilde{k}}{\sigma_c}\left[\int_{\eta_0}^{\eta}
\tilde{\Pi}_2(\eta')\mathrm{d}\eta'- \tilde{k}(\eta - \eta_0)\right]\right)\equiv \nonumber\\&&\equiv \tilde{B}_0(\tilde{k},\eta_0)\exp\left(\int_{\eta_0}^{\eta}\left[\alpha (\eta')\tilde{k} - \tilde{k}^2\beta (\eta')\right]\mathrm{d}\eta'\right) ,
\end{eqnarray}
where $\tilde{B}_0(\tilde{k},\eta_0)=\sqrt{\tilde{k}\tilde{h}(\tilde{k},\eta_0)}$ is the initial seed magnetic field .

Note that for relic neutrinos, which drive the generation of CMF, the sign of neutrino asymmetries is unknown. Thus, meaning the positive $\tilde{\Pi}_2>0$ for the dynamo action in eq.~(\ref{mono}), we should choose $\alpha=|\alpha|$  hence changing negative factor  $F$ in eq.~(\ref{Fdef}) (plotted in figure~\ref{FTmu}) to $|F|$ and substituting $| \xi_{\nu_e} - \xi_{\nu_{\mu}} - \xi_{\nu_{\tau}}|$ in eq.~(\ref{alpha}).

\end{document}